%% file: astroph.tex
\documentclass[usenatbib]{mn2e}

\usepackage{times}
\usepackage{lscape}
\usepackage{natbib}
\usepackage{graphicx}
\input{dra_ssppnewcommands1.tex}

\input{dra_newcommands2.tex}

\input{dra_newcommands3.tex}

\input{dra_newcommands4.tex}

\newcommand{\apjl}{ApJL Letters}
\newcommand{\apjs}{ApJS}
\newcommand{\nat}{Nature}
\newcommand{\aj}{AJ}
\newcommand{\mnras}{MNRAS}
\newcommand{\araa}{ARA\&A}
\newcommand{\aap}{A\&A}
\newcommand{\apj}{ApJ}
\newcommand{\teff}{T_{\mathrm{eff}}}
\newcommand{\feh}{\mathrm{[Fe/H]}}
\newcommand{\alphafe}{[\alpha/\mathrm{Fe}]}
\newcommand{\logg}{\log g}
\newcommand{\teffi}{T_{\mathrm{eff}_i}}
\newcommand{\fehi}{\mathrm{[Fe/H]}_i}

\newcommand{\loggi}{\log g_i}

\newcommand{\temp}{\mathrm{temp}}
\newcommand{\pix}{\mathrm{pix}}
\newcommand{\sky}{\mathrm{sky}}

\newcommand{\var}{\mathrm{Var}}
\newcommand{\los}{\mathrm{los}}
\newcommand{\atm}{\mathrm{atm}}

\usepackage{amsmath}
\newcommand{\angstrom}{\textup{\AA}}

\title[Bayesian analysis of resolved stellar spectra]{Bayesian analysis of resolved stellar spectra: application to MMT/Hectochelle Observations of the Draco dwarf spheroidal}
\author[Walker, Olszewski \& Mateo]{Matthew G. Walker$^{1,2}$\thanks{E-mail:
mgwalker@cmu.edu}, Edward W. Olszewski$^{3}$ and Mario Mateo$^{4}$\thanks{Observations reported here were obtained at the MMT Observatory, a joint facility of the Smithsonian Institution and the University of Arizona.}\\
$^{1}$McWilliams Center for Cosmology, Department of Physics, Carnegie Mellon University,
5000 Forbes Ave., Pittsburgh, PA 15213, United States\\
$^{2}$Harvard-Smithsonian Center for Astrophysics, 60 Garden St.,
Cambridge, MA 02138, United States\\
$^{3}$Steward Observatory, The University of Arizona, 933 N. Cherry
Ave., Tucson, AZ 85721, United States\\
$^{4}$University of Michigan, 311 West Hall, 1085 S. University Ave., Ann Arbor, MI 48109,
United States}
\begin{document}


\pagerange{\pageref{firstpage}--\pageref{lastpage}} \pubyear{2014}

\maketitle

\label{firstpage}

\begin{abstract}
We introduce a Bayesian method for fitting faint, resolved stellar spectra in order to obtain simultaneous estimates of redshift and stellar-atmospheric parameters.  We apply the method to thousands of spectra---covering $5160-5280$ \AA\ at resolution $\mathcal{R}\sim 20,000$---that we have acquired with the MMT/Hectochelle fibre spectrograph for red-giant and horizontal branch candidates along the line of sight to the Milky Way's dwarf spheroidal satellite in Draco.  The observed stars subtend an area of $\sim 4$ deg$^{2}$, extending $\sim 3$ times beyond Draco's nominal `tidal' radius.  For each spectrum we tabulate the first four moments---central value, variance, skewness and kurtosis---of posterior probability distribution functions representing estimates of the following physical parameters: line-of-sight velocity ($v_{\los}$), effective temperature ($\teff$), surface gravity ($\logg$) and metallicity ($\feh$).  After rejecting low-quality measurements, we retain a new sample consisting of \goodobs\ independent observations of \goodstars\ unique stars, including \repeats\ observations for \repeatstars\ stars with (as many as \maxrepeats) repeat observations.  Parameter estimates have median random errors of $\sigma_{v_{\los}}$$=$\medsigv\ km s$^{-1}$, $\sigma_{\teff}$$=$\medsigteff\ K, $\sigma_{\logg}$$=$\medsiglogg\ dex and $\sigma_{\feh}$$=$\medsigfeh\ dex.  Our estimates of physical parameters distinguish $\sim 470$ likely Draco members from interlopers in the Galactic foreground.  
\end{abstract}


\section{Introduction}

The Milky Way's dwarf spheroidal (dSph) satellites include the smallest, oldest, nearest, least luminous and least chemically-enriched galaxies known.  These extreme properties have made dSphs the focus of investigations ranging from galaxy formation to cosmology and particle physics.  Several generations of imaging and spectroscopic surveys have dramatically improved knowledge of dSph luminosity functions, internal structure and kinematics, chemical abundances and star-formation histories.  Thus far every deeper/wider survey has revealed new surprises---examples include the original discovery that dSph gravitational potentials are dominated by dark matter \citep{aaronson83}, discoveries of `ultrafaint' satellites \citep{willman05b,zucker06a,belokurov07}, the discovery of chemo-dynamically independent stellar sub-populations \citep{tolstoy04,battaglia06,battaglia11}, evidence for dwarf-dwarf mergers \citep{amorisco14}, and discoveries of extremely metal-poor stars \citep{kirby08,frebel10b}.  Many investigators are now exploiting the richness of existing data sets in the attempt to piece together a comprehensive understanding of small-scale structure formation within the Galactic neighbourhood.  

\smallskip
The recent explosion of data in this field has been fuelled by rapid advances in instrumentation.  For example, the availability of multi-object, medium- and high-resolution spectrographs at $6-10$m telescopes has increased the yield of a night's observing from a few to a few hundred spectra.  As a result, the number of dSph stars with precise kinematic and chemical abundance measurements has grown by more than an order of magnitude in the past decade \citep{battaglia06,walker09a,kirby10,tollerud11,collins13}.  

\smallskip
In an effort to propel this trend still further and, more specifically, to build samples over the wide fields necessary to study dependences of stellar velocity and metallicity distributions on position, since 2006 we have been using the Hectochelle spectrograph at the 6.5-m Multiple Mirror Telescope (MMT) on Mt. Hopkins, Arizona, to conduct a large-scale stellar-spectroscopic survey of northern dSphs.  During this time we have also developed a Bayesian analysis pipeline that fits a flexible model to each spectrum and delivers multi-dimensional, posterior probability distribution functions (PDFs) for redshift as well as stellar-atmospheric parameters such as effective temperature ($\teff$), surface gravity ($\logg$) and metallicity ($\feh$) .  From the PDFs we derive not only central values and variances corresponding to estimates of each parameter, but also higher-order moments (skewness and kurtosis) that let us gauge quality of a given parameter estimate.  

\smallskip
Here our purpose is three-fold: 1) to describe the acquisition and reduction of our Hectochelle spectra, 2) to introduce our spectral model and the derivation of posterior PDFs and summary statistics for model parameters, and 3) to make publicly available a new data catalogue containing first results from our Hectochelle survey.  As the subject of this initial data release, we choose the Draco dSph (central coordinates $\alpha_{2000}=$17:20:12, $\delta_{2000}=$+57:54:55, distance $D=76\pm 6$ kpc, absolute magnitude $M_V=-8.8\pm 0.3$, halflight radius $r_h=221\pm 19$ pc \citep[and references therein]{mateo98,mcconnachie12}), for which we have acquired, reduced and analysed \obs\  independent Hectochelle spectra for \stars\ unique stars.  From these spectra we obtain high-quality measurements for \goodobs\ observations of \goodstars\ unique stars.

\section{Observations and Data Reduction}

All measurements presented here come from our observations with Hectochelle \citep{szentgyorgyi98,szentgyorgyi06}, a bench-mounted, multi-object, fibre-echelle spectrograph at the 6.5-m MMT on Mt. Hopkins, Arizona.  Hectochelle uses 240 fibres to gather light from individual sources (fibre apertures are $\sim 1\arcsec$) over a field of diameter 1$^{\circ}$.

\subsection{Target Selection}

\begin{figure*}
  \includegraphics[width=6in]{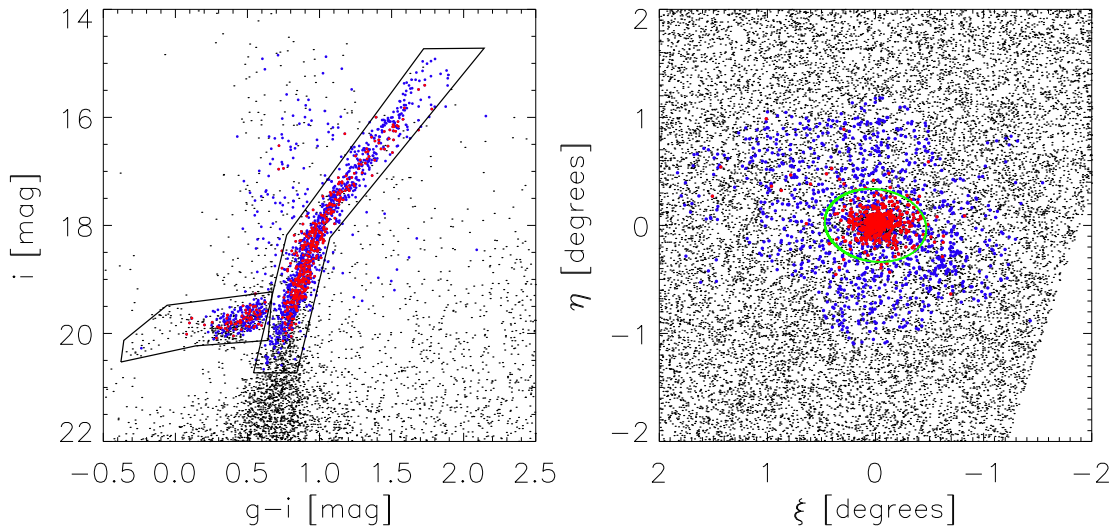}
  \caption{\textit{Left:} Colour-magnitude diagram for point sources toward Draco, from SDSS photometry with extinction corrections applied.  Black points indicate stars projected within $10\arcmin$ of Draco's centre.   Blue and red points indicate all stars---including those at larger angular separations from the centre---for which we obtained Hectochelle spectra.  Red points identify probable Draco members, based on physical parameters estimated from Hectochelle spectra (Section \ref{sec:scatterplots}).    \textit{Right:} Spatial distribution of all stars falling in red-giant and horizontal branch selection boxes shown in the left panel (black points).  Blue and red points again identify all Hectochelle-observed targets, and red points identify probable members.  The solid green ellipse marks the boundary where Draco's surface brightness falls to the background level, as estimated by \citet{ih95}.  }
  \label{fig:dra_rgbhb}
\end{figure*}

The left-hand panel of Figure \ref{fig:dra_rgbhb} shows a colour-magnitude diagram (CMD) for point sources toward Draco, generated with $g$- and $i$-band photometry from the Sloan Digital Sky Survey \citep[SDSS; Data Release 9,][]{ahn12} and corrected for extinction according to the dust maps of \citet{schlegel98}.  Black points indicate stars projected within $10\arcmin$ of Draco's centre.  These centrally-located stars clearly trace Draco's red-giant and horizontal branches (enclosed by polygons in the figure).  Blue and red points indicate all stars---including those at larger angular separations from the centre---for which we obtained Hectochelle spectra.  In general we tried to select spectroscopic targets from red-giant and horizontal branches.  Obvious outliers are associated with early (pre-2009) Hectochelle runs, for which we selected targets, preferentially near Draco's centre, based on uncalibrated photometry taken with the Kitt Peak National Observatory's 0.9m telescope (Mosaic-1 camera) and the Bok 2.3m telescope (90Prime camera).  For later Hectochelle runs we selected targets directly from SDSS photometry.  The right-hand panel of Figure \ref{fig:dra_rgbhb} displays projected spatial distributions for all stars falling in the fiducial red-giant and/or horizontal branch boxes overlaid in the left panel (black points) as well as all Hectochelle-observed targets (blue and red points).  Red points indicate stars for which the Hectochelle data indicate probable Draco membership, based on the criteria discussed in Section \ref{sec:scatterplots}.  

\subsection{Observations} 

Our MMT/Hectochelle observations took place between April 2006 $-$ May 2011, during which we observed 33 unique fibre configurations for Draco.  We observed nine of these fields on two separate occasions, and henceforth we treat these repeat observations independently.  We observed with Hectochelle's `RV31' order-blocking filter, which gives coverage over $5150-5300$ \AA.  The most prominent feature in this region is the Mgb triplet, with rest wavelengths of 5167 \AA, 5173 \AA\ and 5184 \AA; however, the region also includes an assortment of Fe II lines that can provide a direct measure of iron abundance.  Since we bin the detector by factors of 2 and 3 in spectral and spatial dimensions, respectively, our Hectochelle spectra have an effective dispersion of $\sim 0.1$ \AA\ per pixel, or $\sim 0.3$ \AA\ per resolution element, for an effective resolving power of $\mathcal{R}\sim 20,000$.  In a typical fibre configuration, $\sim 200$ fibres are allocated to science targets and $\sim 40$ are allocated to regions of relatively blank sky.  For each configuration, Table \ref{tab:log} lists central coordinates of the field, dates of observation and total exposure times, which are typically the sum over $3-5$ sub-exposures.
\input{dra_log_mnras.tex}

\smallskip
For a given fibre configuration we took Th-Ar arc-lamp exposures both before and after the sequence of science exposures, and we took a quartz-lamp exposure either immediately before or immediately after the science exposures.  For the purpose of calibration and correction for variations in fibre throughput, we also took series of exposures (including Th-Ar and quartz calibrations) during evening and/or morning twilight.  

\subsection{Data Reduction}
\label{subsec:reductions}

Hectochelle's detector consists of two $2048\times 4096$-pixel CCDs, each of which is read out through two amplifiers.  In order to maximize the fibre-packing factor on the CCDs, Hectochelle's fibres are mounted in a zigzag pattern at the slit plane.  As a result, adjacent spectra are offset in both spatial and dispersion directions on the detector.  The entire pattern also curves along the spatial axis, demanding care in identifying and tracing apertures before extraction.  

We use standard IRAF\footnote{IRAF is distributed by the National Optical Astronomy Observatory, which is operated by the Association of Universities for Research in Astronomy, Inc., under cooperative agreement with the National Science Foundation.} routines to process the raw images, to extract one-dimensional spectra and to estimate the wavelength solution for each spectrum obtained in each science exposure.  The entire procedure is similar to those previously described by \citet{walker07a} and \citet{mateo08}, except that here we also propagate the variance associated with the count in each pixel of the raw images.  At the outset, for every science frame (i.e., the image obtained in an individual science exposure), we generate a corresponding variance frame in which the value assigned to a given pixel is
\begin{equation}
  \var(\pix)=C(\pix)G+R^2,
  \label{eq:pixelvariance}
\end{equation}
where $C(\pix)$ is the count in analogue-to-digital units (ADU), $G\approx 1.03$ e$^{-}$/ADU is the gain of the Hectochelle detector and $R=3$ e$^{-}$ is read noise.  In order to propagate variances, we process variance frames in the same way that we process their corresponding science frames (see below).  Where we rescale count levels in a given science  frame---in order, e.g., to correct for variability in fibre throughput--- we rescale variances by (the square of) the same factor.  Where we combine spectra via addition or subtraction (e.g., to combine sub-exposures or to subtract sky background), we compute the combined variance as the sum of the variances associated with the pixels contributing to the sum or difference.  

\smallskip
For a given frame we use the IRAF package CCDPROC to perform overscan and bias corrections independently for each amplifier.  For each of the two CCDs, we then combine arrays from the two amplifiers to form a continuous image.  Next we use the IRAF package APALL to identify the locations and shapes of spectral apertures, and then to extract one-dimensional spectra for science, quartz and Th-Ar exposures and associated variance frames.  We initially identify aperture locations and trace patterns in the relatively bright quartz frames.  Fixing the relative locations and shapes of apertures according to the quartz frames, we allow the entire aperture pattern to shift globally in order to provide the best match to the spectra detected in the corresponding science frames; typical shifts are $\la 0.01$ pixels; we apply exactly the same shift to define apertures and traces for the Th-Ar frame.  For science, quartz, ThAr and associated variance frames, we extract one-dimensional spectra from each aperture by combining (adding) counts from pixels along the axis perpendicular to the dispersion direction.  

\smallskip
Next we use the extracted quartz spectra to adjust for differences in fibre throughput and pixel sensitivity.  We fit the quartz spectrum in each aperture with a $10^{\rm th}$-order cubic spline, then divide each continuum fit by the mean value over all pixels in the entire frame.  Then we divide each science and quartz spectrum by the normalized fit to the quartz continuum obtained in the same aperture, thereby adjusting for throughput differences.  Finally we divide the science spectra by the throughput-corrected quartzes, thereby correcting for differences in pixel sensitivity.  

\smallskip
Next we estimate wavelength solutions, $\lambda(\pix)$.  For each extracted ThAr spectrum, we use the IRAF package IDENTIFY to fit a $9^{\rm th}$-order cubic spline to the centroids of between $\sim 35-40$ identified emission lines of known wavelength.  Residuals of these fits typically have root mean square (rms) scatter $\sim 0.006$ \AA, or $\sim 0.3$ km s$^{-1}$, similar to the minimum velocity error we determine from observations of solar twilight spectra (Section \ref{subsec:twilights}).  We assign the same aperture-dependent wavelength solutions to the corresponding science frames.  Except for extraction from 2D to 1D, each spectrum retains the sampling native to the detector, such that wavelength solutions $\lambda(\pix)$ generally differ from one spectrum to another and have non-uniform $\Delta\lambda/\Delta\pix$ even within the same spectrum.

\smallskip
Next we identify and reject pixels affected by cosmic rays.  We fit the continuum of each wavelength-calibrated science spectrum with a $5^{\rm th}$-order polynomial, iteratively rejecting pixels with values that either exceed the fit by more than 3.5 times the rms of residuals, or are smaller than the fit by more than 1.75 times the rms of residuals (the smaller value of the latter tolerance excludes absorption features from the fit).  After ten iterations, we replace the variances of all high outliers with large values ($10^{100}$), effectively removing their influence on all subsequent analysis.  

\subsection{Sky Subtraction}
Next we estimate and subtract sky background from all extracted, sensitivity-corrected and wavelength-calibrated spectra.  For the observed spectral range ($5150-5300$ \AA), telluric emission is negligible and the dominant source of sky background is scattered sunlight.  Most of our spectra were obtained in near-dark conditions, such that sky backgrounds are typically small.  Nevertheless, for all spectra we follow the sky-subtraction procedure of \citet{koposov11}.  Specifically, for a given frame we interpolate each of the $N_{\sky}\sim 30$ individual (one-dimensional, throughput-corrected) sky spectra onto a common grid with constant spacing $\Delta\lambda'/\Delta\pix'=0.01$\AA\ (i.e., oversampled by a factor of $\sim 10$ with respect to the original sampling).  For each discrete wavelength of the oversampled sky spectrum we record the median count level and estimate the variance as $2.198\pi(2N_{\sky})^{-1}(\mathrm{MAD}^2)$, where $\mathrm{MAD}$ is the median absolute deviation \citep{koposov11,rousseeuw93}.  We then smooth\footnote{We use the IDL function `INTERPOLATE' for the interpolation and smoothing steps.} the resulting spectrum of median sky level (and associated variance) back onto the real, irregularly-sampled wavelength solution that is unique to a given target spectrum, so that we can then subtract the median sky spectrum from the target spectrum, pixel by pixel.

\begin{figure*}
  \includegraphics[width=\textwidth]{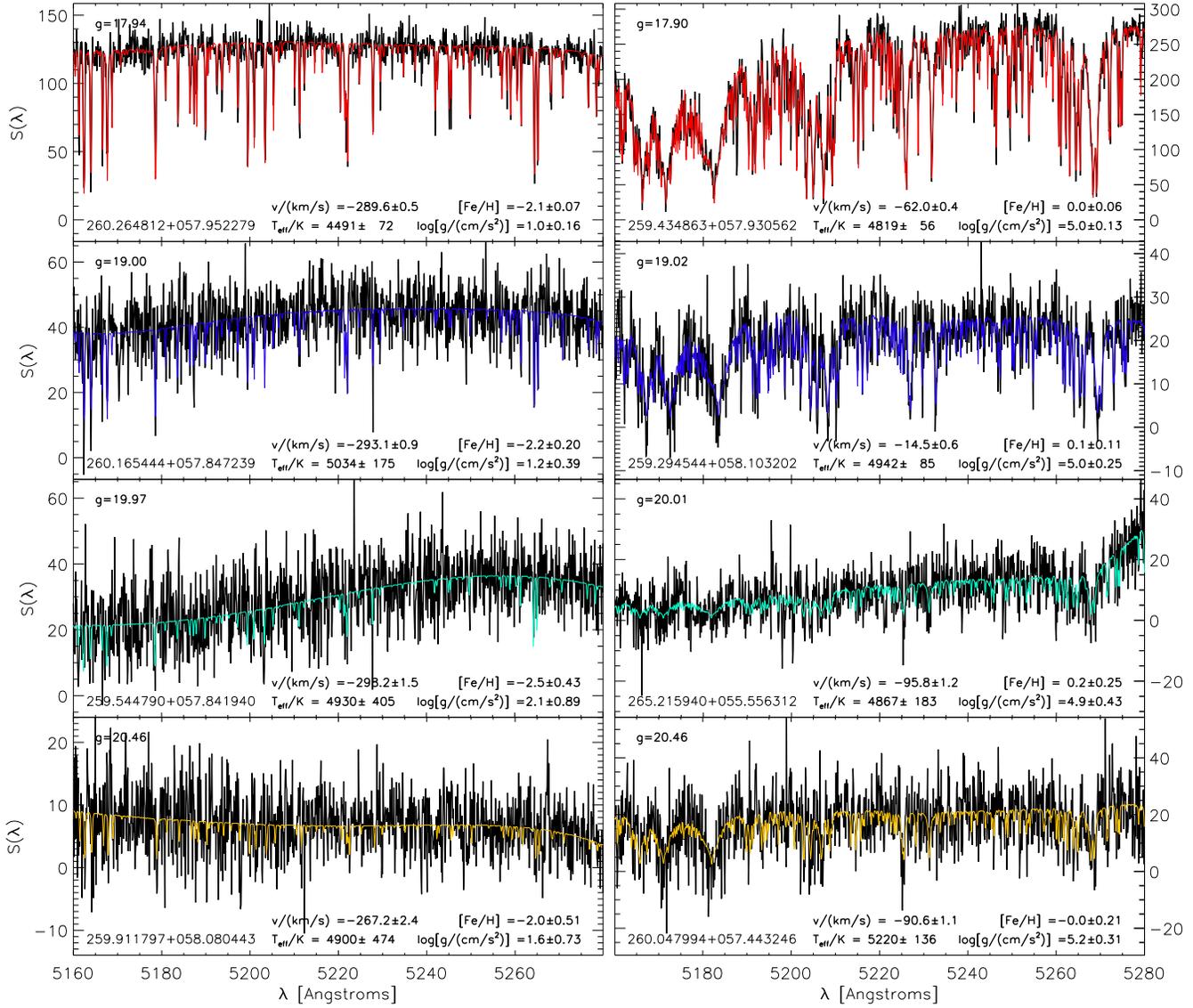}
  \caption{\scriptsize Sky-subtracted Hectochelle spectra (black) for probable members of Draco (left panels) and probable foreground dwarf stars (right panels) spanning magnitudes $18\la g\la 20.5$.  Overplotted with coloured curves are best-fitting model spectra synthesized from the spectral library of \citet{delaverny12}.  Text lists equatorial coordinates, SDSS $g$-band magnitude, and our estimates of $v_{\los}$, $\teff$, $\logg$, $\feh$, $\alphafe$.}
  \label{fig:dra_spectra}
\end{figure*}

\smallskip
Finally, we combine sub-exposures (which have identical apertures and wavelength solutions) by taking the inverse-variance-weighted mean at each pixel of the sky-subtracted spectra.  Figure \ref{fig:dra_spectra} displays examples of the resulting Hectochelle spectra for science targets spanning the magnitude range $18\la g\la 20.5$.

\section{Analysis of Spectra}
\label{sec:analysis}
We fit each sky-subtracted spectrum using a flexible model that adapts easily to different instrumental setups and scientific goals.  The model is built from `template' spectra that can, in principle, be real spectra of standard stars or synthetic spectra calculated for specific stellar-atmospheric models.   Requirements for templates are 1) they must have high signal-to-noise ratios, 2) their resolution must be similar to or higher than that of the science spectra, and 3) their parameters of interest  (e.g., stellar-atmospheric parameters) must be known.  

\subsection{Spectral Model}
\label{subsec:model}
Following \citet{koleva09} and \citet{koposov11}, we consider a spectral model of the form
\begin{equation}
  M(\lambda)=P_l(\lambda)T\biggl (\lambda\biggl [1+\frac{Q_m(\lambda)+v_{\los}}{c}\biggr ]\biggr ),
  \label{eq:model}
\end{equation}
where $c$ is the speed of light.  The order-$l$ polynomial
\begin{eqnarray}
  P_l(\lambda)\equiv p_0+p_1\biggl [\frac{\lambda-\lambda_0}{\lambda_s}\biggr ]+p_2\biggl [\frac{(\lambda-\lambda_0)}{\lambda_s}\biggr ]^2\nonumber\\
+\ldots+p_l\biggl [\frac{(\lambda-\lambda_0)}{\lambda_s}\biggr ]^l
  \label{eq:polynomial1}
\end{eqnarray}
gives shape to the continuum and $T(\lambda[1+\frac{Q_m(\lambda)+v_{\los}}{c}])$ is a continuum-normalized template that is redshifted according to a given 1) line-of-sight velocity, $v_{\rm los}$, and 2) order-$m$ polynomial, 
\begin{eqnarray}
  Q_m(\lambda)\equiv q_0+q_1\biggl [\frac{\lambda-\lambda_0}{\lambda_s}\biggr ]+q_2\biggl [\frac{(\lambda-\lambda_0)}{\lambda_s}\biggr ]^2\nonumber\\
  +...+q_m\biggl [\frac{(\lambda-\lambda_0)}{\lambda_s}\biggr ]^m.
  \label{eq:polynomial2}
\end{eqnarray}
The latter, wavelength-dependent redshift accounts for linear and higher-order systematic differences between wavelength solutions of target and template spectra.  Notice that the zero$^{\rm th}$-order term, $q_0$, is degenerate with $v_{\los}$; therefore we assume in all cases that $q_0=0$, leaving our velocity estimates susceptible to a uniform error in zero point that we can estimate using spectra from stars of known velocity (e.g., the solar spectra taken at twilight, see Section \ref{subsec:twilights}).

\subsubsection{Templates}
\label{subsubsec:templates}

Lacking Hectochelle spectra spanning a sufficient range of standard stars, we generate template spectra using a library of synthetic spectra that has generously been provided by Young Sun Lee (private communication, 2013), and has previously been used to estimate stellar-atmospheric parameters as part of the SEGUE Stellar Parameter Pipeline \citep[`SSPP' hereafter,]{lee08a,lee08b}.  The SSPP library contains rest-frame, continuum-normalized, stellar spectra computed over a grid of atmospheric parameters spanning $4000\leq T_{\mathrm{eff}}/\mathrm{K}\leq 10000$ in effective temperature (with spacing $\Delta T_{\mathrm{eff}}/\mathrm{K}=250$), $0\leq\log_{10}[g/(\mathrm{cm/s^2})]\leq 5$ in surface gravity ($\Delta\log_{10}[g/(\mathrm{cm/s^2})]=0.25$ dex) and $-5\leq \mathrm{[Fe/H]}\leq +1$ in metallicity ($\Delta\mathrm{[Fe/H]}=0.25$ dex).  While the library spectra are calculated for a range in the abundance ratio of $\alpha$ elements to iron, this ratio is not independent of metallicity.  Rather, $\alphafe=+0.4$ for library spectra with $\feh < -1$, $\alphafe$ decreases linearly as metallicity increases from $-1\leq \feh< 0$, and then $\alphafe=0$ for $\feh\geq 0$.  The synthetic library spectra are calculated over the range $3000-10000$ \AA\ at resolution $0.01$ \AA\ /pixel, which we degrade to $0.05$ \AA\ per pixel---twice as fine as our Hectochelle spectra---in order to reduce computational costs.  

\smallskip
We denote as $L_0(\lambda,\vec{\theta}_{\atm})$ the original library spectrum corresponding to the `vector' of stellar-atmospheric parameters $\vec{\theta}_{\atm}\equiv (\teff,\logg,\feh)$.  In order to account for the instrumental line-spread function (LSF), we smooth each library spectrum using a Gaussian kernel, such that the smoothed version of the library spectrum is
\begin{equation}
  L(\lambda,\vec{\theta}_{\atm},h_0)=\frac{\displaystyle \sum_{i=1}^{N_{\lambda,\mathrm{temp}}}L_0(\lambda_i,\vec{\theta}_{\atm})K_1\biggl(\frac{\lambda_i-\lambda}{h_0} \biggr)}{\displaystyle\sum_{i=1}^{N_{\lambda,\mathrm{temp}}}K_1\biggl(\frac{\lambda_i-\lambda}{h_0} \biggr)},
  \label{eq:smooth}
\end{equation}
where $N_{\lambda,\mathrm{temp}}$ is the number of pixels in the library spectrum and the kernel is
\begin{equation}
  K_1\biggl (\frac{\lambda_i-\lambda}{h_0}\biggr )\equiv \exp \biggr[ -\frac{1}{2}\frac{(\lambda_i-\lambda)^2}{h_0^2}\biggr]. 
  \label{eq:kernel}
\end{equation}
In order to reduce computational cost further during the subsequent fitting procedure, we initially generate three different versions of each library spectrum using smoothing bandwidths $h_0/$\AA$=0.05, 0.075, 0.1$.   These values were determined empirically to span the full range of bandwidths required to fit the library's version of the solar spectrum ($\teff=5750$ K, $\log_{10}[g/(\mathrm{cm/s^2})]=4.5$, $\feh=0$) to the solar twilight spectra acquired through each fibre during our Hectochelle runs.

Finally, in order to let the spectral model vary continuously despite the library's coarse gridding in stellar-atmospheric parameter space and the three discrete values for the initial smoothing bandwidth, we apply a second smoothing kernel over the entire collection of smoothed library spectra.  Specifically, for any choice of stellar-atmospheric parameters, $\vec{\theta}_{\atm}$, and smoothing bandwidth, $h_0$, we obtain a unique template,
\begin{equation}
  T(\lambda)=\frac{\displaystyle\sum_{i=1}^{N_L}L(\lambda,\vec{\theta}_{\atm_i}, h_{0_i})K_2\biggl (\frac{\vec{\theta}_{\atm_i}-\vec{\theta}_{\atm}}{\vec{h}_{\atm}},\frac{h_{0_i}-h_0}{h_h}\biggr )}{\displaystyle\sum_{i=1}^{N_L}K_2\biggl (\frac{\vec{\theta}_{\atm_i}-\vec{\theta}_{\atm}}{\vec{h}_{\atm}},\frac{h_{0_i}-h_0}{h_h}\biggr )},
  \label{eq:template}
\end{equation}
where $N_L=25275$ is the number of library spectra (including three smoothed versions for each of the 8425 original library spectra) and the kernel is
\begin{eqnarray}
  K_2\biggl (\frac{\vec{\theta}_{\atm_i}-\vec{\theta}_{\atm}}{\vec{h}_{\atm}},\frac{h_{0_i}-h_0}{h_h}\biggr )\equiv\hspace{1.2in}\nonumber\\ 
  \exp\biggl [ -\frac{1}{2}\biggl ( \frac{(\teffi-\teff)^2}{h^2_{\teff}}+ \frac{(\loggi-\logg)^2}{h^2_{\logg}}\nonumber\\
  +\frac{(\fehi-\feh)^2}{h^2_{\feh}}+\frac{(h_{0_i}-h_{0})^2}{h^2_h}\biggr ) \biggr ].
  \label{eq:templatesmooth}
\end{eqnarray}
We set smoothing bandwidths equal to the grid spacing in each dimension: $h_{\teff}=250$ K, $h_{\logg}=0.25$ dex, $h_{\feh}=0.25$ dex and $h_h=0.025$ \AA.  Compared with alternative interpolation schemes that we tried, this smoothing procedure gives posterior probability distributions that are more Gaussian and less likely to cluster near the library's grid points. 

\subsection{Likelihood Function and Free Parameters}
\label{subsec:likelihood}
Given the spectral model, $M(\lambda)$, we assume that the observed, sky-subtracted spectrum, $S(\lambda)$, has likelihood
\begin{eqnarray}
  \mathcal{L}\bigl ( S(\lambda) | \vec{\theta},s_1,s_2 \bigr )= \hspace{2.2in}\nonumber \\ \displaystyle\prod_{i=1}^{N_{\lambda}}  \frac{1}{\sqrt{2\pi(s_1\var[S(\lambda_i)]+s_2^2)}}\exp\biggl [-\frac{1}{2} \frac{\bigl ( S(\lambda_i)-M(\lambda_i) \bigr )^2}{s_1\var[S(\lambda_i)]+s_2^2}\biggr ],\nonumber\\   
  \label{eq:likelihood}
\end{eqnarray}
where we have introduced two additional nuisance parameters, $s_1$ and $s_2$, that rescale and add in quadrature, respectively, to observational errors in order to account for cases of mis-estimated noise.  In practice, the value of $M(\lambda_i)$ that we use in Equation \ref{eq:likelihood} is the linear interpolation, at observed wavelength $\lambda_i$, of the discrete model we calculate from Equation \ref{eq:model}.  This interpolation is necessary because a given template spectrum, $T(\lambda)$, retains the discrete wavelength sampling of the synthetic library, which generally differs from those of the observed spectra.

\smallskip
We fit to the Hectochelle spectra over the region $5160 \leq (\lambda/\angstrom$) $\leq 5280$, which gives $N_{\lambda}\sim 1150$.  For the templates we consider rest-frame wavelengths $5150 \leq \lambda/$\AA\ $\leq 5290$, giving $N_{\lambda,\temp}=1400$ and allowing redshifts corresponding to the range $-575 \leq v_{\los}/($ km s$^{-1})\leq +575$.  

\smallskip
In order to define the polynomials in Equations \ref{eq:polynomial1} and \ref{eq:polynomial2}, we choose $l=5$ and $m=2$, respectively.  These choices give sufficient flexibility to fit the continuum shape and to apply low-order corrections to the wavelength solution.  We adopt scale parameters $\lambda_0=5220$ \AA\  and $\lambda_s=60$ \AA, such that $-1 \leq (\lambda-\lambda_0)/\lambda_s \leq +1$ over the entire range considered in the fits. 

\smallskip
With these choices the spectral model, $M(\lambda)$, is fully specified by a vector of 13 free parameters:  
\begin{eqnarray}
  \vec{\theta}
  =(v_{\los},\teff,\logg,\feh,\hspace{1.73in}\nonumber\\
  p_0,p_1,p_2,p_3,p_4,p_5,q_1,q_2,h_0 )\hspace{0.3in}.
  \label{eq:params}
\end{eqnarray}
The first four have physical meaning and the rest are nuisance parameters that specify coefficients of polynomials, smoothing bandwidth and adjustments to observational errors.  Table \ref{tab:priors} lists all parameters and identifies the adopted priors, all of which are uniform over a specified range of values and zero outside that range. 
\begin{table*}
  \scriptsize
  \centering
  \begin{minipage}{140mm}
    \caption{Free parameters and priors}
    \begin{tabular}{@{}lllllll@{}}
      \hline
      parameter& prior& description\\
      \hline
      $v_{\los}/(\mathrm{km/s})$&uniform between $-500,+500$&line-of-sight velocity\\
      $\teff/K$&uniform between $4000,8000$&effective temperature\\
      $\log_{10}[g/(\mathrm{cm}/\mathrm{s}^2)]$&uniform between $0,5$& surface gravity\\
      $\feh$&uniform between $-5,+1$& iron abundance\\
      $p_0$&uniform between$^{a}$ $-\max[S(\lambda)],+\max[S(\lambda)]$&polynomial coefficient (continuum; eq \ref{eq:polynomial1})\\
      $p_1$&uniform between $-\max[S(\lambda)],+\max[S(\lambda)]$&polynomial coefficient (continuum; eq \ref{eq:polynomial1})\\
      $p_2$&uniform between $-\max[S(\lambda)],+\max[S(\lambda)]$&polynomial coefficient (continuum; eq \ref{eq:polynomial1})\\
      $p_3$&uniform between $-\max[S(\lambda)],+\max[S(\lambda)]$&polynomial coefficient (continuum; eq \ref{eq:polynomial1})\\
      $p_4$&uniform between $-\max[S(\lambda)],+\max[S(\lambda)]$&polynomial coefficient (continuum; eq \ref{eq:polynomial1})\\
      $p_5$&uniform between $-\max[S(\lambda)],+\max[S(\lambda)]$&polynomial coefficient (continuum; eq \ref{eq:polynomial1})\\
      $q_1/(\mathrm{km/s})$&uniform between $-10,+10$&polynomial coefficient (wavelength solution; eq. \ref{eq:polynomial2})\\
      $q_2/(\mathrm{km/s})$&uniform between $-10,+10$&polynomial coefficient (wavelength solution; eq. \ref{eq:polynomial2})\\
      $h_0/$\AA$]$&uniform between $0.05,0.10$&broadens line-spread function (eq. \ref{eq:smooth})\\
      $\log_{10}s_1$&uniform between $-1,+6$&rescales observational errors (eq. \ref{eq:likelihood}) \\
      $\log_{10}s_2$&uniform between $-2,+2$&adds to
      observational errors (eq. \ref{eq:likelihood}) \\
      \hline
    \end{tabular}
    \\
    $^{a}$ $\max[S(\lambda)]$ is the maximum value (discounting
    pixels flagged as cosmic rays) of the sky-subtracted spectrum.
  \end{minipage}
  \label{tab:priors}
\end{table*}

\subsection{Parameter Estimation}
\label{subsec:fitting}
\smallskip
From Bayes' theorem, given the observed spectrum, $S(\lambda)$, the model has posterior PDF
\begin{equation}
  p\bigl ( \vec{\theta},s_1,s_2 | S(\lambda)\bigr ) = \frac{ \mathcal{L}\bigl ( S(\lambda)|\vec{\theta},s_1,s_2\bigr ) p ( \vec{\theta},s_1,s_2)}{p\bigl ( S(\lambda)\bigr )},
  \label{eq:bayes}
\end{equation}
where $\mathcal{L} (S(\lambda)|\vec{\theta},s_1,s_2)$ is the likelihood from Equation \ref{eq:likelihood}, $p(\vec{\theta},s_1,s_2)$ is the prior and 
\begin{equation}
  p\bigl ( S(\lambda)\bigr ) \equiv \displaystyle\int \mathcal{L}\bigl ( S(\lambda) | \vec{\theta},s_1,s_2 \bigr ) p(\vec{\theta},s_1,s_2) d\vec{\theta}ds_1ds_2
  \label{eq:evidence}
\end{equation}
is the marginal likelihood, or `evidence'.  

In order to evaluate the posterior PDF we must scan the 15-dimensional parameter space.  For this task we use the software package MultiNest\footnote{available at ccpforge.cse.rl.ac.uk/gf/project/multinest} \citep{feroz08,feroz09}.  MultiNest implements a nested-sampling Monte Carlo algorithm \citep{skilling04} that is designed to calculate the evidence (Equation \ref{eq:evidence}) and simultaneously to sample the posterior PDF (Equation \ref{eq:bayes}).  Briefly, the algorithm proceeds by evolving a set of (in our case, 2000) `live points' that initially are distributed throughout the parameter volume according to the prior.  Then, in each iteration, the point with lowest likelihood is replaced by a new point that is drawn randomly from the prior, subject to the condition that the new point must have likelihood larger than that of the point being replaced.  As the number of iterations increases, the set of live points becomes confined within regions of higher likelihood and smaller volume, allowing the evidence to be computed with increasing accuracy.  The algorithm terminates when the accuracy reaches a specified threshold (we adopt a tolerance of 0.5 in log-evidence, with sampling efficiency of 0.8, as recommended by MultiNest's README file).  \citet{feroz08} and \citet{feroz09} demonstrate that MultiNest performs well even when the posterior is multi-modal and has strong curving degeneracies, circumstances that can present problems for standard Markov-Chain Monte Carlo (MCMC) techniques.  In our own experiments, MultiNest consistently required fewer likelihood evaluations than did the widely-used Metropolis-Hastings algorithm \citep{metropolis53,hastings70}, which was more prone to becoming trapped in the local minima generated by the presence of multiple absorption features.  In order to fit a single Hectochelle spectrum, we typically require a few$\times 10^5$ evaluations of Equation \ref{eq:likelihood}.

\section{Results}
\label{sec:results}

Plotted over each example Hectochelle spectrum in Figure \ref{fig:dra_spectra} is the best-fitting model spectrum, as calculated from the most likely (according to Equation \ref{eq:likelihood}) point from the posterior sample returned by MultiNest.  Text within each panel indicates estimates of physical parameters $v_{\los}$ (transformed to the solar rest frame), $\teff$, $\logg$ and $\feh$.  Based on these estimates, spectra in left-hand panels correspond to probable members of Draco, with velocities near Draco's mean of $\sim -290$ km s$^{-1}$, low surface gravities ($\logg \la 3.5$) characteristic of red giants, and low metallicities ($\feh\la -2$).  Spectra in right-hand panels correspond to probable contaminants in the Galactic foreground, which are likely to be G dwarfs with stronger surface gravities ($\logg\ga 4$) and metallicities ($\feh\ga -1$).   Visually the foreground dwarfs have stronger absorption lines than do the bona fide red giants, and our spectral modelling is clearly sensitive to these differences.

\begin{figure*}
  \includegraphics[width=\textwidth]{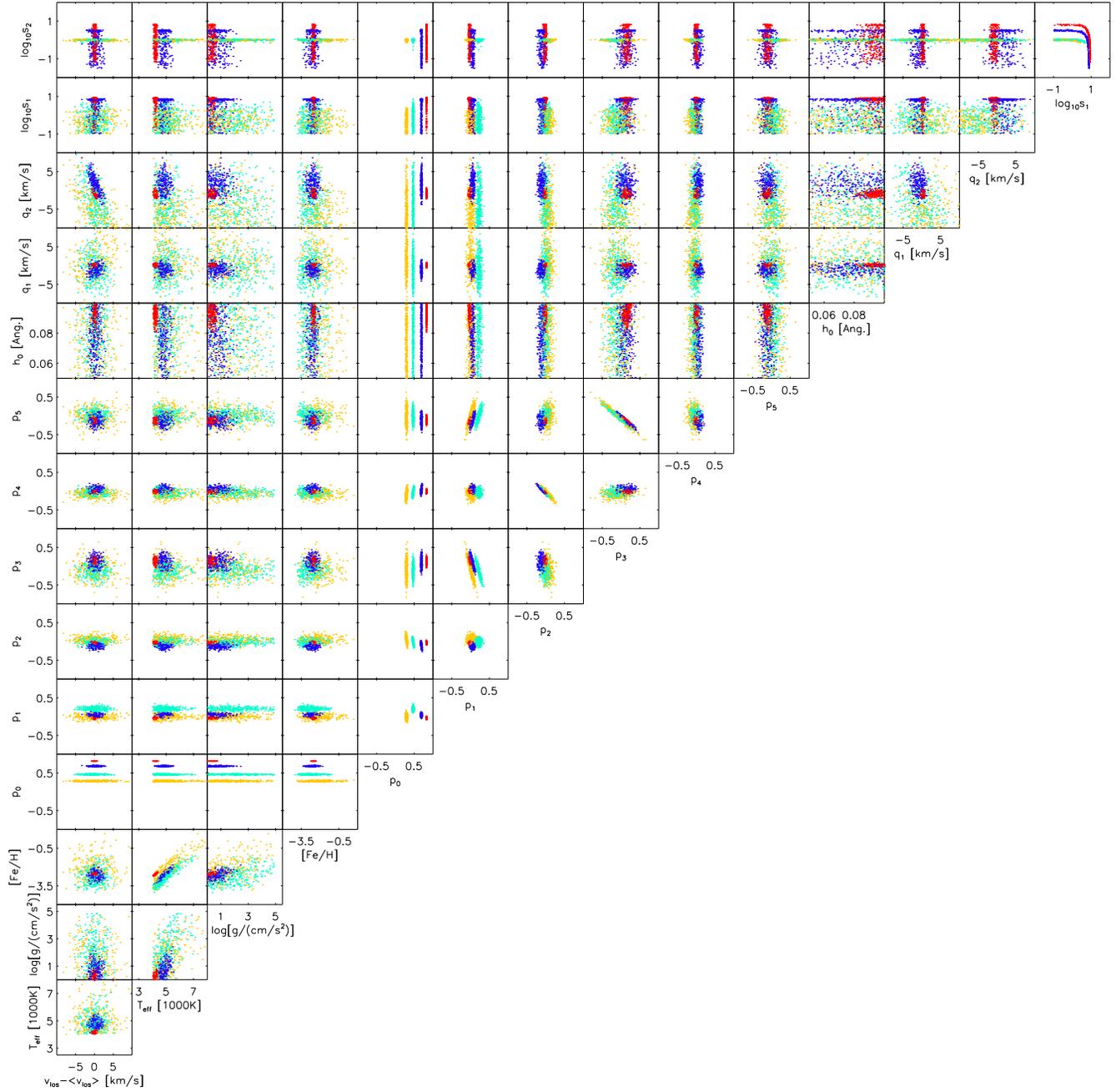}
  \caption{\scriptsize Correlations among model parameters, from samplings of the posterior distribution function obtained in fits to the spectra of probable Draco members shown in Figure \ref{fig:dra_spectra}.  Colours correspond to those used for best-fitting spectra in Figure \ref{fig:dra_spectra}.} 
  \label{fig:dra_params}
\end{figure*}

\smallskip
For the four spectra corresponding to probable Draco members in Figure \ref{fig:dra_spectra}, Figure \ref{fig:dra_params} shows samples from the posterior PDF, as returned by MultiNest.  These plots reveal various degeneracies among model parameters.  Of the five physical parameters, $\teff$, $\logg$ and $\feh$ all show correlations.  These degeneracies make sense because a relatively weak absorption feature can result from relatively low metal abundance, lower opacity in the relatively diffuse atmosphere that exists in a weaker gravitational field and/or relatively high degrees of ionization in hotter atmospheres.  On the other hand, lack of correlation between stellar-atmospheric parameters and $v_{\los}$ implies that redshifts estimated from full spectra are relatively insensitive to the spectral types of templates.  Regarding nuisance parameters, Figure \ref{fig:dra_params} reveals correlations among polynomial coefficients that describe the continuum.  The curving degeneracy between nuisance parameters $s_1$ and $s_2$ indicates that the model prefers to adjust variance spectra either by rescaling or by adding a constant value, but not both.  

\subsection{Posterior Probability Distribution Functions}
\label{subsec:pdfs}

For a given spectrum, MultiNest's sampling of the posterior PDF lets us evaluate not only the degeneracies discussed above, but also moments of the marginalized, 1-D posterior PDF for each model parameter.  Given $N$ random draws from the posterior PDF, then for a given parameter, $X$, we calculate moments of the error distribution as follows.  The first moment is the mean, $\overline{X}\equiv N^{-1}\sum_{i=1}^Nx_i$, which we take to be the central value.  The second moment is the variance, $\sigma_X^2\equiv (N-1)^{-1}\sum_{i=1}^N(x_i-\overline{X})^2$, which we take to be the square of the $1\sigma$ credibility interval (for a Gaussian error distribution, the interval $\overline{X}\pm \sigma_X$ encloses $68\%$ of the integrated probability).  The third moment is skewness, $S\equiv N^{-1}\sum_{i=1}^N[(x_i-\overline{X})/\sqrt{\sigma_X^2}]^3$, which equals zero for a symmetric distribution.  The fourth moment is kurtosis, $K\equiv N^{-1}\sum_{i=1}^N[(x_i-\overline{X})/\sqrt{\sigma^2_X}]^4$, which distinguishes Gaussian distributions ($K=3$) from `leptokurtic' ones with sharper peaks and fatter tails ($K>3$) and `platykurtic' ones with broader peaks and weaker tails ($K<3$).  

\begin{figure*}
  \includegraphics[width=\textwidth]{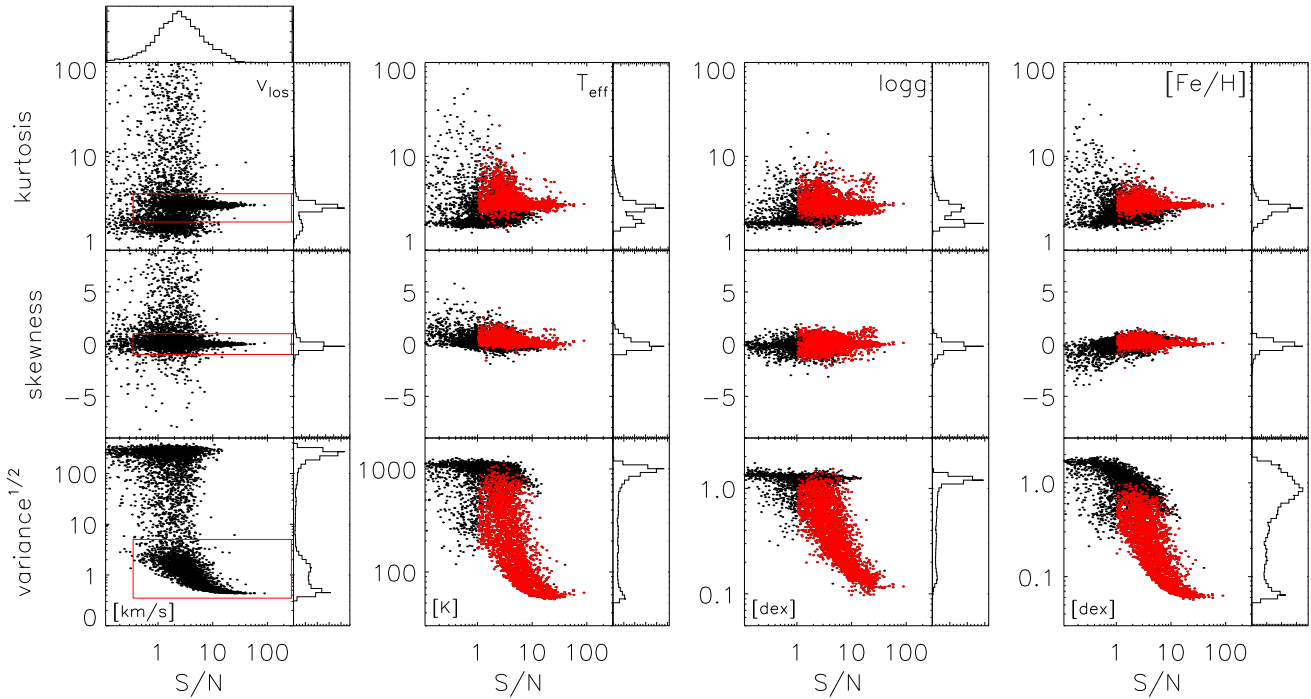}
  \caption{\scriptsize Moments of marginalized, 1D posterior probability distribution functions (PDFs) of model parameters versus median signal-to-noise ratio per pixel (a resolution element is $\sim 3$ pixels), for \obs\ independent Hectochelle observations of \stars\ unique stars.  Histograms show marginalized distributions for each moment as well as S/N.  Red rectangles enclose regions where posterior PDFs of velocity estimates are near-Gaussian ($-1\leq$ skewness$\leq +1$, $2\leq$ kurtosis$\leq 4$) and have $\sigma_{v_{\los}}\leq 5$ km s$^{-1}$.   Observations associated with these estimates are coloured red in other panels; subsequent analysis and data tables refer only to these observations.}
  \label{fig:dra_diagnostics}
\end{figure*}

\smallskip
For each of our \obs\ independent Hectochelle spectra of \stars\ unique stars, scatterplots in Figure \ref{fig:dra_diagnostics} show how moments of the PDFs for physical parameters depend on the median signal-to-noise ratio per pixel.  Each point in a given panel corresponds to a single spectrum.  As S/N grows, the PDFs tend to become more Gaussian ($S\sim 0$, $K\sim 3$) and have smaller variance; thus the distributions shown in Figure \ref{fig:dra_diagnostics} provide objective criteria for quality control.  In the velocity panels, we identify a cluster of \goodobs\ measurements that have $\sigma_{v_{\los}}\leq 5$ km s$^{-1}$, $-1\leq S\leq +1$ and $2\leq K\leq 4$.  The relatively small variance and near-Gaussianity of these posterior PDFs imply tight constraints for which the $68\%$ credibility interval can be specified with reasonable accuracy by $\overline{v}_{\los}\pm \sigma_{v_{\los}}$.  Therefore, in all subsequent analysis and in the data tables presented below, we include only observations for which moments of posterior PDFs for $v_{\los}$ fall within the ranges specified above (red squares in the velocity panels of Figure \ref{fig:dra_diagnostics}).  Because some analyses of these data will require only velocity information, here we do not make further cuts based on moments of posterior PDFs for other physical parameters; however, the data tables list all four moments of the posterior PDF for each parameter, so users can apply any such cuts as needed.  Furthermore, all spectra and samples from posterior PDFs are available for download from the following web address: http://www.andrew.cmu.edu/user/mgwalker/hectochelle.

\subsection{Accuracy, Precision and Dependence on Templates}
\label{subsec:twilights}

\begin{figure*}
  \includegraphics[width=\textwidth]{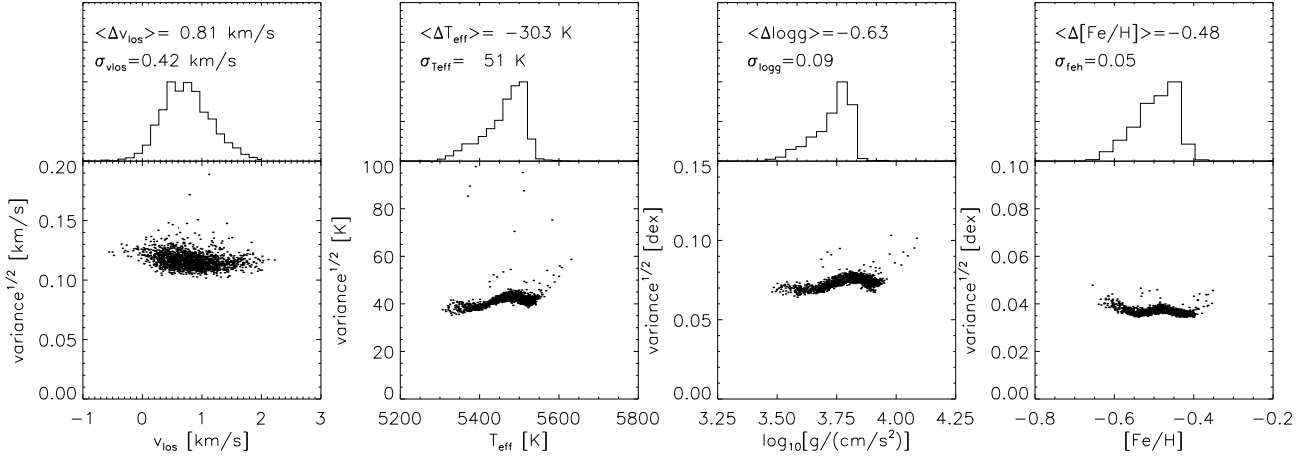}
  \caption{\scriptsize Estimates of physical parameters for $1715$ solar spectra acquired with Hectochelle in morning/evening twilight during our Draco runs.  Scatterplots show first and (square root of) second moments of posterior probabillity distribution functions; histograms show distributions of first moments.  Text indicates 1) mean offset of first moments (means) with respect to known solar values ($v_{\los,\odot}=0$, $T_{\mathrm{eff},\odot}=5778$ K, $\log_{10}[g/\mathrm{cm/s^2})]=4.44$, $\feh=0$), and 2) square root of second moments (variances) of the distribution of first moments.  We subtract the former from first moments of PDFs obtained for our science spectra, and we add the latter (in quadrature) to square roots of second moments of PDFs obtained for our science spectra; these adjustments are included in the values reported in Table \ref{tab:dra_table1}.  }
  \label{fig:dra_twilights}
\end{figure*}

In order to gauge both accuracy and precision of our estimates of physical parameters, we examine fits---using the same procedure described above---to $1715$ high-S/N (tens of thousands of counts per pixel) solar spectra acquired with Hectochelle in morning/evening twilight during our Draco runs.  For each physical parameter, scatterplots in Figure \ref{fig:dra_twilights} show first and (square root of) second moments of the posterior PDFs.  Histograms show distributions of the first moments, which let us quantify empirically 1) mean offsets with respect to known solar values ($v_{\los,\odot}=0$, $T_{\mathrm{eff},\odot}=5778$ K, $\log_{10}[g_{\odot}$/(cm s$^{-2})]=4.44$, $\feh_{\odot}=0$) and 2) scatter.  For stellar-atmospheric parameters the empirical scatter is consistent with second moments of the posterior PDFs returned by MultiNest.   For $v_{\los}$ the empirical scatter of $\sim 0.4$ km s$^{-1}$ is several times larger than the scatter implied by the posterior PDFs, the larger value reflects the typical rms of residuals in our wavelength calibrations (Section \ref{subsec:reductions}).  For each physical parameter, $X$, Table \ref{tab:offsets} lists the mean offset, $\langle \Delta\overline{X} \rangle \equiv \langle \overline{X}-X_{\odot}\rangle$, and the standard deviation, $\sigma_{\overline{X}}$.  

\begin{table}
  \scriptsize
  \centering
  \begin{minipage}{140mm}
    \caption{Fits to twilight spectra for three synthetic libraries}
    \begin{tabular}{@{}llll@{}}
      \hline
      &SSPP$^{a}$&C05$^{b}$&AMBRE$^{c}$\\
      \hline
      $\langle \overline{v_{\los}}-v_{\los,\odot}\rangle$&\ssppdeltav\ km s$^{-1}$&\coelhodeltav\ km s$^{-1}$&$0.76$ km s$^{-1}$\\
      $\langle \overline{\teff}-T_{\mathrm{eff},\odot}\rangle$&\ssppdeltateff\ K&\coelhodeltateff\ K&$22$ K\\
      $\langle \overline{\logg}-\logg_{\odot}\rangle$&\ssppdeltalogg\ dex&\coelhodeltalogg\ dex&$0.08$ dex\\
      $\langle \overline{\feh}-\feh_{\odot}\rangle$&\ssppdeltafeh\ dex&\coelhodeltafeh\ dex&$-0.34$ dex\\
      \\
      $\sigma_{\overline{v_{\los}}} $&\ssppsigmav\ km s$^{-1}$&\coelhosigmav\ km s$^{-1}$&$0.43$ km s$^{-1}$ \\
      $\sigma_{\overline{\teff}} $&\ssppsigmateff\ K&\coelhosigmateff\ K&$43$ K\\
      $\sigma_{\overline{\logg}} $&\ssppsigmalogg\ dex&\coelhosigmalogg\ dex&$0.10$ dex \\
      $\sigma_{\overline{\feh}} $&\ssppsigmafeh\ dex&\coelhosigmafeh\ dex&$0.03$ dex  \\
      \hline
    \end{tabular}
    \\
    \raggedright
    $^{a}$ library of \citet{lee08a,lee08b}\\
    $^{b}$ library of \citet{coelho05}\\
    $^{c}$ library of \citet{delaverny12}\\
  \end{minipage}
  \label{tab:offsets}
\end{table}

\smallskip
We use these values to adjust raw parameter estimates for science targets.  For each parameter we subtract the empirical offset from the first moments obtained from the posterior PDFs.  Then we add the empirical scatter, in quadrature, to the (square roots of) second moments for all Draco observations.  All subsequent discussion and analysis (and data tables) refer to results that have been adjusted in this way.  For the \goodobs\ observations that pass the quality-control criteria described in Section \ref{subsec:pdfs}, our estimates of physical parameters have median (minimum, maximum) random errors of $\sigma_{v_{\los}}$$=$\medsigv\ (\minsigv, \maxsigv) km s$^{-1}$, $\sigma_{\teff}$$=$\medsigteff\ (\minsigteff, \maxsigteff) K, $\sigma_{\logg}$$=$\medsiglogg\ (\minsiglogg, \maxsiglogg) dex and $\sigma_{\feh}$$=$\medsigfeh\ (\minsigfeh, \maxsigfeh) dex.

\smallskip

We note that the offsets of our estimates from solar values are unique to our use of the SSPP library of synthetic spectra.  We have repeated our entire fitting procedure, for both science and twilight spectra, using the alternative libraries of \citet[][`C05' hereafter]{coelho05} and \citet[][`AMBRE' hereafter]{delaverny12}.  Table \ref{tab:offsets} lists solar offsets obtained for each physical parameter with each library.  Those obtained with the C05 library are negligible in the sense that they are smaller than the random errors indicated by the posterior PDFs for the twilight spectra.  The AMBRE library gives a significant offset from the solar value only for metallicity, where $\langle \overline{\feh}-\feh_{\odot}\rangle=-0.34$ dex.  

\smallskip
Despite seemingly-better performances of the C05 and AMBRE libraries, we report results only from our fits with the SSPP library.  The irregularity and coarseness (spacing of $\Delta\feh=0.5$ dex for $\feh<-1$ and $\Delta\feh=1$ dex for $\feh<-3$) of the AMBRE grid introduce relatively large and asymmetric smoothing errors near the parameter space most relevant to Draco.  The C05 library contains no models with $\feh<-2.5$ and the resulting fits to many of our Draco spectra fail to give a lower bound on $\feh$.  Reassuringly, when we compare raw estimates obtained from higher-metallicity ($\feh\sim -2.0$) Draco spectra, for which fits with both the SSPP and C05 library provide lower bounds on $\feh$, we recover the same relative offsets---for each physical parameter---obtained in fits to solar spectra.  This agreement suggests that the SSPP offsets listed in Table \ref{tab:offsets} do not depend strongly on the stellar-atmospheric parameters themselves and thus can be applied to our Draco spectra.  Indeed, after we apply these offsets our measurements for Draco stars come into good agreement with those of previous studies (Section \ref{sec:previous}). 

\subsection{Repeatability}
\label{subsec:repeatability}

\begin{figure*}
  \includegraphics[width=\textwidth]{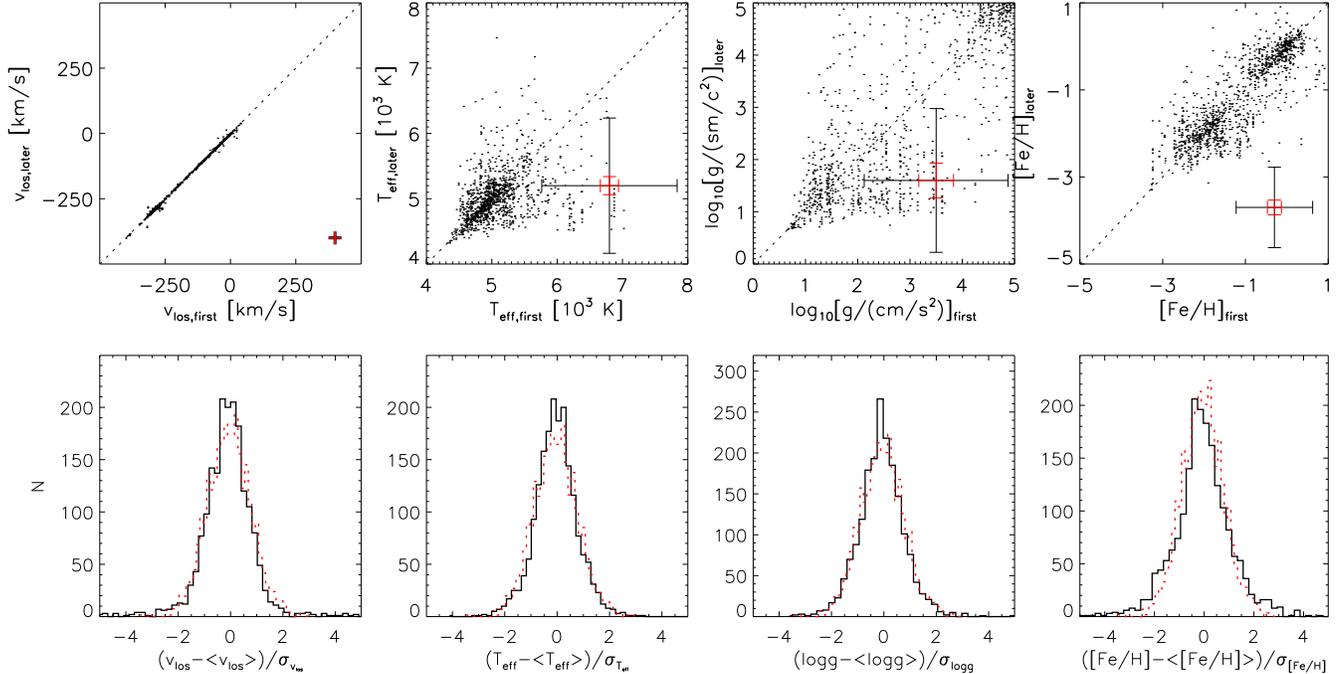}
  \caption{\scriptsize Repeatability of parameter estimates, from \repeats\ independent observations of \repeatstars\ unique stars.  For each (physical) model parameter, scatterplots in the top row indicate estimates obtained from later observations against those obtained from the first observation.  Straight lines indicate 1:1 relations, and errorbars in the lower-right corner of each panel indicate median (red) and maximum (black) credibility intervals associated with the plotted points.  Histograms in the bottom row indicate distributions of deviations with respect to the weighted mean, $\langle \overline{X}\rangle \equiv \sum_{i=1}^N (\overline{X}_i/\sigma^2_{X_i})/\sum_{i=1}^{N_{\rm obs}}\sigma_{X_i}^{-2}$, normalized by credibility intervals, $\sigma_{X_i}$, for $N_{\rm obs}$ independent measurements of each star with repeat observations.  Dotted red histograms indicate distributions obtained from artificial data having the same numbers of observations for each star as in the real data, but with `observed' values scattered only according to measurement errors.} 
  \label{fig:dra_repeats}
\end{figure*}

\begin{figure}
  \includegraphics[width=3in]{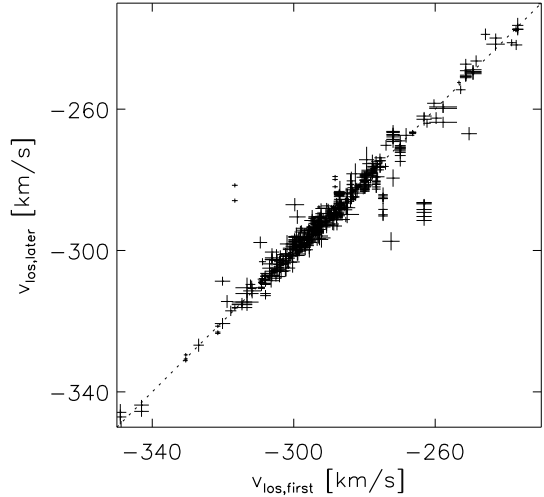}
  \caption{\scriptsize Same as top-left panel of Figure \ref{fig:dra_repeats}, but zoomed in on the region populated by Draco members.} 
  \label{fig:dra_repeats_zoom}
\end{figure}

After discarding observations for which posterior PDFs for $v_{\los}$ fail to meet the criteria described in Section \ref{subsec:pdfs}, there remain \goodobs\ independent observations of \goodstars\ stars.  These include \repeats\ observations of \repeatstars\ stars with multiple observations passing the criteria described in Section \ref{subsec:pdfs}, with as many as \maxrepeats\ such observations for a single star.  In order to examine repeatability of our parameter estimation, we compare results from multiple, independent measurements of a given star.  For all science targets with repeat observations, top panels in Figure \ref{fig:dra_repeats} compare estimates derived from the first observation to those derived in each subsequent observation (Figure \ref{fig:dra_repeats_zoom} zooms in on the velocity range of Draco members).  With few exceptions, points scatter around 1:1 relations according to the variances of the posterior PDFs.  

More quantitatively, bottom panels in Figure \ref{fig:dra_repeats} show distributions of deviations with respect to inverse-variance-weighted means---$\langle \overline{X}\rangle \equiv \sum_{i=1}^N (\overline{X}_i/\sigma^2_{X_i})/\sum_{i=1}^{N_{\rm obs}}(1/\sigma_{X_i}^{2})$---normalized by the propagated error.  We do not expect these distributions to be Gaussian, because the calculation of weighted means introduces correlations among deviations that correspond to observations of the same star.  Therefore, instead of comparing the observed distributions to Gaussians, we compare them to distributions (red histograms in Figure \ref{fig:dra_repeats}) that we obtain from artificial data that have the same number of observations for each of the same number of stars, but with scatter introduced only by observational errors.  Comparisons between the observed and artificial distributions show excellent agreement for $v_{\los}$, $\teff$ and $\logg$, indicating that the variances of the PDFs returned by MultiNest are indeed reliable estimates of 68\% credibility intervals.  For $\feh$, the observed distribution has thicker tails than the artificial one, indicating that the variances of the PDFs returned by MultiNest correspond to slightly underestimated credibility intervals.  

\section{Data Tables}
\label{sec:data}

For all \goodobs\ individual observations that pass the quality-control criteria described in Section \ref{subsec:pdfs}, Table \ref{tab:dra_table1} lists equatorial coordinates, heliocentric Julian date at the beginning of the first sub-exposure, median signal-to-noise ratio per pixel, and moments estimated from posterior PDFs for physical parameters $v_{\los}$ (transformed to the heliocentric rest frame), $\teff$, $\logg$, $\feh$ and $\alphafe$.  For all model parameters, the reported interval is $\overline{X}\pm \sigma_X$---i.e., the central value is the first moment (mean) of the posterior PDF, and the credible interval is given by the square root of the second moment (square root of the variance).  Third (skewness) and fourth (kurtosis) moments are listed parenthetically as superscripts above the credible interval.  Table \ref{tab:dra_table2} then lists one set of estimates for each of the \goodstars\ unique stars.  For stars with multiple independent measurements, Table \ref{tab:dra_table2} lists the weighted mean values for model parameters.  Users that require stricter quality-control criteria (e.g., cuts on moments of posterior PDFs for stellar-atmospheric parameters) are advised to re-compute these weighted means after refining Table \ref{tab:dra_table1}.  

\begin{table*}
\scriptsize
\caption{Hectochelle data from individual observations$^{a}$}
\begin{tabular}{@{}lrccccccccccccccccccccc@{}}
\hline
$\alpha_{2000}$&$\delta_{2000}$&HJD&S/N$^{b}$&$\overline{v_{\rm los}}$&$\overline{T_{\rm eff}}$&$\overline{\log_{10}g}$&$\overline{\mathrm{[Fe/H]}}$\\

[hh:mm:ss]&[$^{\circ}$:$\arcmin$:$\arcsec$]&[days]&&[km s$^{-1}$]$^{c}$&[K]&[dex]$^{d}$&[dex]\\
\hline
17:20:40.31&+57:56:18.1&$2455331.74$&$  3.1$&$-293.4\pm 0.8^{(-0.1,3.3)}$&$4806\pm 199^{( 0.1, 3.3)}$&$ 0.9\pm 0.2^{( 1.2, 4.4)}$&$-1.63\pm0.25^{( 0.0, 3.0)}$\\
17:21:17.89&+57:56:21.2&$2455331.74$&$  4.4$&$-292.0\pm 0.7^{( 0.1,3.0)}$&$4613\pm 152^{( 0.1, 2.8)}$&$ 1.1\pm 0.3^{( 0.5, 2.9)}$&$-1.94\pm0.18^{( 0.3, 3.0)}$\\
17:23:26.43&+58:08:10.7&$2455331.74$&$  1.6$&$ -37.3\pm 1.2^{( 0.1,3.3)}$&$4779\pm 201^{( 0.0, 2.7)}$&$ 4.0\pm 0.6^{( 0.1, 2.9)}$&$ 1.01\pm0.29^{(-0.4, 2.5)}$\\
 & &$2455706.84$&$  7.6$&$ -34.5\pm 0.5^{(-0.0,3.0)}$&$5089\pm  65^{( 0.1, 3.0)}$&$ 5.0\pm 0.2^{( 0.1, 3.2)}$&$ 0.43\pm0.08^{( 0.0, 3.1)}$\\
 & &$2455707.75$&$ 11.5$&$ -34.4\pm 0.5^{( 0.0,2.9)}$&$5124\pm  64^{( 0.1, 3.0)}$&$ 4.8\pm 0.1^{( 0.1, 3.0)}$&$ 0.11\pm0.07^{(-0.1, 3.0)}$\\
 & &$2455710.91$&$  6.7$&$ -33.9\pm 0.5^{(-0.0,3.0)}$&$5008\pm  69^{( 0.1, 3.1)}$&$ 4.8\pm 0.2^{( 0.1, 3.0)}$&$ 0.25\pm0.08^{( 0.0, 3.0)}$\\
17:20:27.38&+57:56:52.4&$2455331.74$&$  8.6$&$-294.5\pm 0.6^{( 0.0,3.0)}$&$4620\pm 106^{(-0.2, 3.2)}$&$ 0.8\pm 0.2^{( 1.0, 3.8)}$&$-2.23\pm0.13^{( 0.0, 3.0)}$\\
17:21:44.98&+58:10:50.5&$2455331.74$&$  2.6$&$ -18.7\pm 1.1^{(-0.1,2.9)}$&$4918\pm 118^{(-0.1, 2.9)}$&$ 5.0\pm 0.3^{(-0.4, 2.7)}$&$-0.03\pm0.17^{(-0.1, 3.3)}$\\
 & &$2455708.93$&$  6.6$&$ -17.1\pm 0.5^{(-0.1,2.9)}$&$4885\pm  69^{(-0.1, 3.0)}$&$ 5.3\pm 0.2^{(-0.2, 2.7)}$&$-0.25\pm0.08^{(-0.0, 3.0)}$\\
 & &$2455712.87$&$  6.8$&$ -17.1\pm 0.5^{( 0.0,3.1)}$&$4842\pm  65^{( 0.0, 2.9)}$&$ 4.9\pm 0.2^{(-0.0, 3.1)}$&$-0.31\pm0.08^{(-0.1, 3.0)}$\\
17:20:47.88&+58:08:14.3&$2455706.77$&$  5.4$&$ -50.6\pm 0.5^{(-0.0,3.0)}$&$5332\pm  83^{( 0.1, 3.0)}$&$ 5.0\pm 0.2^{( 0.1, 3.1)}$&$ 0.08\pm0.09^{( 0.0, 3.1)}$\\
 & &$2455707.75$&$  6.2$&$ -50.0\pm 0.6^{( 0.0,3.0)}$&$5320\pm  90^{(-0.1, 2.9)}$&$ 5.2\pm 0.2^{(-0.2, 2.6)}$&$-0.25\pm0.10^{(-0.1, 3.0)}$\\
 & &$2455708.93$&$  3.0$&$ -50.2\pm 0.7^{( 0.1,3.1)}$&$5009\pm 110^{( 0.1, 3.2)}$&$ 4.2\pm 0.3^{(-0.1, 3.1)}$&$ 0.04\pm0.13^{(-0.1, 3.1)}$\\
 & &$2455712.87$&$  3.2$&$ -50.8\pm 0.7^{( 0.0,3.0)}$&$5143\pm 105^{( 0.1, 3.0)}$&$ 4.8\pm 0.3^{(-0.0, 3.0)}$&$ 0.19\pm0.13^{( 0.1, 3.1)}$\\
17:21:18.40&+58:22:53.6&$2455331.74$&$  0.9$&$-131.6\pm 1.5^{(-0.3,3.2)}$&$4939\pm 265^{(-0.2, 2.5)}$&$ 4.7\pm 0.7^{(-0.7, 2.9)}$&$ 0.90\pm0.34^{(-0.4, 2.6)}$\\
 & &$2455708.86$&$  9.3$&$-132.5\pm 0.5^{(-0.0,3.1)}$&$5101\pm  69^{( 0.1, 3.1)}$&$ 5.0\pm 0.2^{( 0.1, 3.3)}$&$-0.51\pm0.08^{(-0.1, 3.1)}$\\
17:21:47.10&+58:21:51.3&$2455331.74$&$  4.1$&$  -7.8\pm 0.7^{( 0.0,2.9)}$&$4744\pm  85^{(-0.1, 3.1)}$&$ 5.0\pm 0.3^{( 0.0, 3.0)}$&$ 0.17\pm0.11^{( 0.0, 3.0)}$\\
17:20:18.69&+58:08:40.7&$2455331.74$&$  2.8$&$  -6.6\pm 0.9^{(-0.0,3.2)}$&$4959\pm 128^{( 0.1, 3.0)}$&$ 4.5\pm 0.4^{( 0.1, 2.9)}$&$-0.03\pm0.18^{(-0.2, 3.3)}$\\
 & &$2455706.84$&$ 10.5$&$  -7.2\pm 0.5^{(-0.1,3.3)}$&$5173\pm  59^{(-0.2, 3.0)}$&$ 5.5\pm 0.1^{(-0.5, 2.7)}$&$-0.25\pm0.07^{(-0.1, 3.1)}$\\
 & &$2455708.93$&$  7.3$&$  -6.8\pm 0.5^{( 0.2,3.2)}$&$5088\pm  64^{(-0.1, 2.9)}$&$ 5.4\pm 0.2^{(-0.3, 2.5)}$&$-0.30\pm0.08^{(-0.0, 3.0)}$\\
 & &$2455712.87$&$  7.3$&$  -7.2\pm 0.5^{(-0.0,2.9)}$&$5150\pm  62^{(-0.3, 3.0)}$&$ 5.5\pm 0.2^{(-0.7, 3.0)}$&$-0.23\pm0.08^{(-0.0, 3.0)}$\\
\hline
\end{tabular}
\\
\raggedright
$^{a}$see electronic edition for complete data table.\\
$^{b}$median signal-to-noise ratio per pixel\\
$^{c}$line-of-sight velocity in the heliocentric rest frame\\
$^{d}$units of $g$ are cm s$^{-2}$\\
\label{tab:dra_table1}
\end{table*}

\begin{table*}
\scriptsize
\caption{Hectochelle data with weighted means for stars with multiple observations$^{a}$}
\begin{tabular}{@{}lrccccccccccccccccccccc@{}}
\hline
$\alpha_{2000}$&$\delta_{2000}$&$g_{\rm SDSS}$&$i_{\rm SDSS}$&$\langle \overline{v_{\rm los}}\rangle $&$\langle \overline{T_{\rm eff}}\rangle $&$\langle \overline{\log_{10}g}\rangle $&$\langle \overline{\mathrm{[Fe/H]}}\rangle $&$N_{\mathrm{obs}}$\\

[hh:mm:ss]&[$^{\circ}$:$\arcmin$:$\arcsec$]&[mag]$^{b}$&[mag]$^{b}$&[km s$^{-1}$]$^{c}$&[K]&[dex]$^{d}$&[dex]\\
\hline
17:20:40.31&+57:56:18.1& 18.94& 17.88&$-293.4\pm 0.8$&$4806\pm 199$&$ 0.9\pm 0.2$&$-1.63\pm0.25$&$  1$\\
17:21:17.89&+57:56:21.2& 18.86& 17.78&$-292.0\pm 0.7$&$4613\pm 152$&$ 1.1\pm 0.3$&$-1.94\pm0.18$&$  1$\\
17:23:26.43&+58:08:10.7& 18.81& 17.76&$ -34.4\pm  0.3$&$5066\pm   37$&$ 4.9\pm  0.1$&$ 0.27\pm 0.04$&$  4$&\\
17:20:27.38&+57:56:52.4& 18.07& 16.79&$-294.5\pm 0.6$&$4620\pm 106$&$ 0.8\pm 0.2$&$-2.23\pm0.13$&$  1$\\
17:21:44.98&+58:10:50.5& 18.65& 17.50&$ -17.2\pm  0.3$&$4870\pm   44$&$ 5.0\pm  0.1$&$-0.26\pm 0.05$&$  3$&\\
17:20:47.88&+58:08:14.3& 19.51& 18.62&$ -50.4\pm  0.3$&$5230\pm   47$&$ 4.9\pm  0.1$&$ 0.00\pm 0.05$&$  4$&\\
17:21:18.40&+58:22:53.6& 19.04& 18.08&$-132.4\pm  0.5$&$5090\pm   67$&$ 5.0\pm  0.2$&$-0.43\pm 0.08$&$  2$&\\
17:21:47.10&+58:21:51.3& 18.10& 16.79&$  -7.8\pm 0.7$&$4744\pm  85$&$ 5.0\pm 0.3$&$ 0.17\pm0.11$&$  1$\\
17:20:18.69&+58:08:40.7& 18.66& 17.63&$  -7.0\pm  0.3$&$5126\pm   34$&$ 5.4\pm  0.1$&$-0.24\pm 0.04$&$  4$&\\
\hline
\end{tabular}
\\
\raggedright
$^{a}$see electronic edition for complete data table.\\
$^{b}$from SDSS photometry, corrected for extinction\\
$^{c}$line-of-sight velocity in the heliocentric rest frame\\
$^{d}$units of $g$ are cm s$^{-2}$\\
\label{tab:dra_table2}
\end{table*}

\section{Comparisons with Previous Work}
\label{sec:previous}

Spectroscopic surveys including more than $\sim 100$ Draco stars have previously been published by \citet[][`A95' hereafter]{armandroff95}, \citet[][`K02' hereafter]{kleyna02} and \citet[][`K10' hereafter]{kirby10}.  In their line-of-sight velocity survey, A95 used the KPNO 4-m telescope and Hydra multifibre spectrograph (spectral range $4720 - 5460$ \AA, resolution $\sim 1.2$ \AA\ per pixel) to measure 259 velocities, with median uncertainty $\sim 4$ km s$^{-1}$, for 159 red giant candidates in Draco.  A95 then merged their sample with that of \citet{edo95}, who used the `old'\footnote{At the time, the MMT consisted of six 1.8-m mirrors, for an effective diameter of $4.5$ m.} MMT's original echelle spectrograph and photon-counting reticon (spectral range $5160 - 5213$ \AA, resolution $\sim 0.2$ \AA\ per pixel) to measure 69 velocities, with median uncertainty $\sim 1.7$ km s$^{-1}$, for 24 probable Draco members.  K02 used the 4.3-m William Herschel Telescope and WYFFOS fibre spectrograph (spectral range $8200-8800$ \AA, resolution $\sim 0.6$ \AA\ per pixel) to measure 159 velocities, with mean precision $\sim 2$ km s$^{-1}$, for 159 probable Draco members (K02 do not tabulate results for an additional 43 probable non-members).  K10 used the 10-m Keck-II Telescope and DEIMOS multi-slit spectrograph (spectral range $6400-9000$ \AA, resolution $\sim 1.2$ \AA\ per resolution element) to measure velocities\footnote{While K10 did not publish velocity measurements, Evan Kirby (private communication) has made their measurements available for the purpose of comparing to our results.} and multi-element abundances (including [Fe/H], [Mg/Fe], [Ca/Fe] and [Ti/Fe]) for 297 probable Draco members.  

\begin{figure*}
  \includegraphics[width=\textwidth]{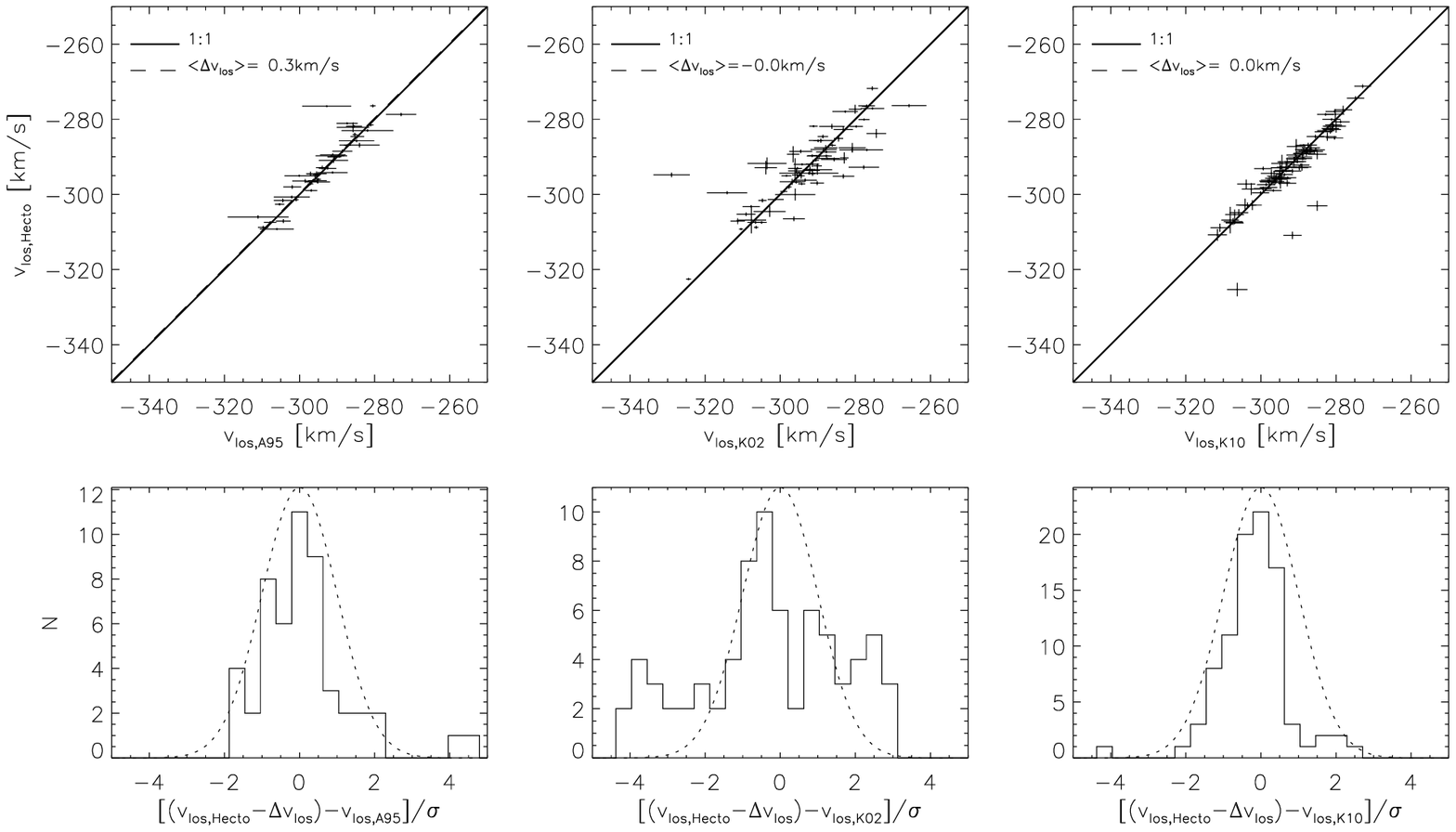}
  \caption{\scriptsize Comparison of Hectochelle velocities to published and/or measured values for stars observed in previous studies.  \textit{Top:} Hectochelle velocities against those published by \citet[][left]{armandroff95} and \citet[][middle]{kleyna02}, and against unpublished velocities generously provided by \citet[][right, private communication]{kirby10}.  Solid lines indicate 1:1 relation; dashed lines indicate the (constant) offset about which scatter is minimized.  \textit{Bottom:} Histograms for velocity differences---normalized by $\sigma\equiv \sqrt{\sigma^2_{V,\mathrm{Hecto}}+\sigma^2_{V,\mathrm{other}}}$, where $\sigma_{V,\mathrm{Hecto}}$ and $\sigma_{V,\mathrm{other}}$ are reported errors for the two measurements---after applying offsets indicated by the dashed lines in top panels (these offsets are \textit{not} applied to the data reported in Tables \ref{tab:dra_table1}-\ref{tab:dra_table2}).  Dotted curves indicate Gaussian distributions with unit variance.}
  \label{fig:dra_previous2}
\end{figure*}

\smallskip
Figure \ref{fig:dra_previous2} compares our velocity estimates to those of A95, K02 and K10, for which our sample has in common 52, 76 and 109 stars, respectively.  Top panels plot our measurements\footnote{For stars with multiple Hectochelle observations, we combine independent estimates using the weighted mean.} directly against the earlier ones; to each of these relations we fit straight lines with slopes of unity, effectively allowing for systematic offsets in zero point.  With respect to the A95, K02 and K10 surveys, these fits suggest zero-point offsets of $\Delta v_{\los}\equiv v_{\rm los, Hecto}-v_{\rm los, other}=$\ssppdeltavarmandroff\ km s$^{-1}$, \ssppdeltavkleyna\ km s$^{-1}$ and \ssppdeltavkirby\ km s$^{-1}$, respectively.  

\smallskip
Bottom panels in Figure \ref{fig:dra_previous2} show histograms of velocity differences, $(v_{\los, \mathrm{Hecto}}-\Delta v_{\los})-v_{\los,\mathrm{other}}$, normalized by combined errors, $\sigma\equiv\sqrt{\sigma^2_{v_{\los,\mathrm{Hecto}}}+\sigma^2_{v_{\los,\mathrm{other}}}}$.  While our median random velocity error is less than half those reported by any of the previous studies, our results stand in generally good agreement with those of A95 and K10, with the possible exceptions of a few outliers that are plausibly explained by intrinsic velocity variability (e.g., binary stars, \citealt{edo96,hargreaves96b}).  However, velocity deviations with respect to the K02 catalogue tend to be larger than can be attributed to reported errors.  We note that the K02 catalogue does not include repeat velocity measurements, making it difficult to check accuracy of their reported errors.  

\begin{figure*}
  \includegraphics[width=\textwidth]{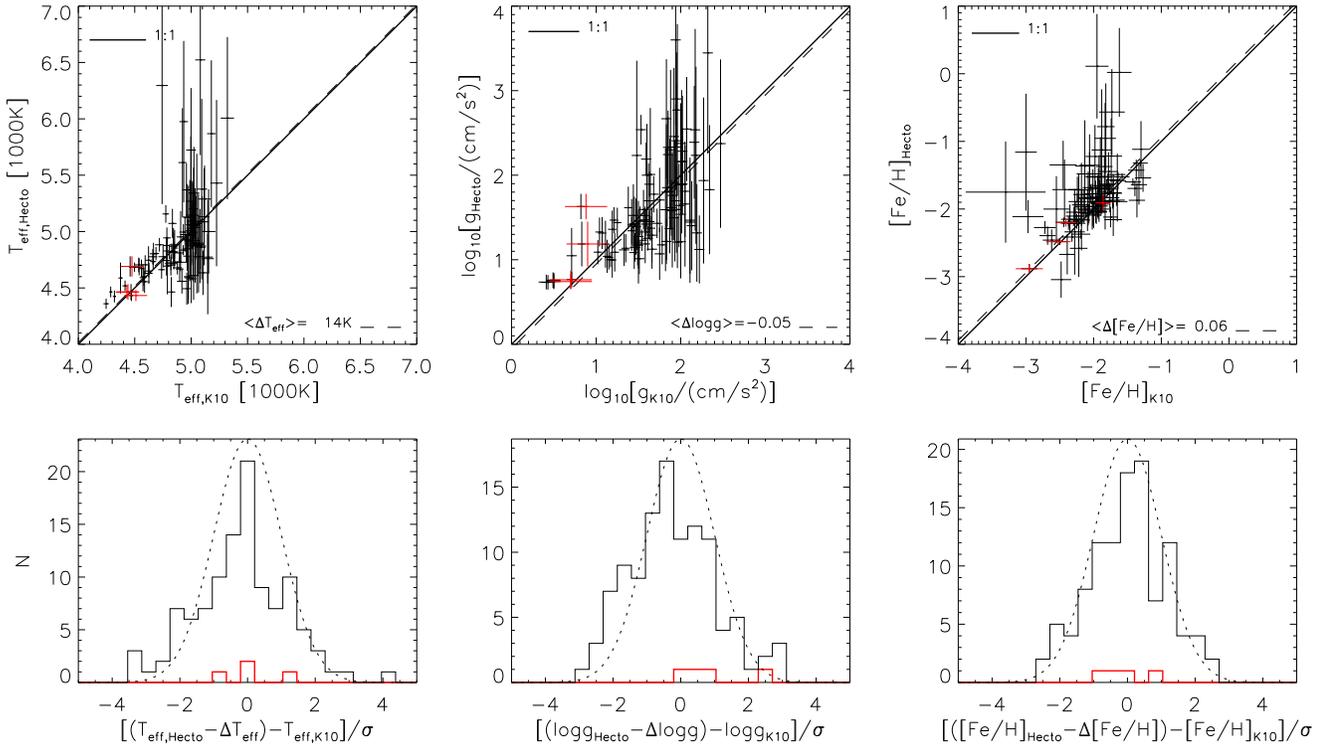}
  \caption{\scriptsize Same as Figure \ref{fig:dra_previous2}, but for comparisons to stellar-atmospheric parameters published by \citet[][black points]{kirby10}.  Red points and histograms show comparisons for three stars with previously-published estimates from Keck/HIRES spectra \citep{shetrone01,fulbright04,cohen09}. }  
  \label{fig:dra_previous1}
\end{figure*}

\smallskip
Figure \ref{fig:dra_previous1} compares our estimates of stellar-atmospheric parameters to those of K10, who estimate $\feh$ from their Keck spectra but determine $\teff$ from a combination of photometry and spectroscopy and determine $\logg$ from photometry alone.  K10 have already demonstrated good agreement between their estimates of stellar-atmospheric parameters and those obtained from high-resolution Keck/HIRES spectra \citep{shetrone01,fulbright04,cohen09}.  In turn, having applied the offsets described in Section \ref{subsec:twilights}, we find good agreement between our estimates and those of K10.  Our fits of straight lines with slopes of unity indicate offsets of $\langle\Delta\teff\rangle\equiv \langle T_{\mathrm{eff,Hecto}}-T_{\mathrm{eff,K10}}\rangle=$\ssppdeltateffkirby\ K, $\langle\Delta\logg\rangle\equiv \langle \log_{10}[g_{\mathrm{Hecto}}]-\log_{10}[g_{\mathrm{K10}}]=$\ssppdeltaloggkirby\ dex (cgs units) and $\langle\Delta\feh\rangle\equiv \langle \mathrm{[Fe/H]}_{\rm Hecto}-\mathrm{[Fe/H]}_{\rm K10}\rangle=$ \ssppdeltafehkirby\ dex; each of these offsets is smaller than the smallest random error among accepted measurements in our data set.  Histograms in lower panels of Figure \ref{fig:dra_previous1} indicate that deviations between our estimates and those of K10 are distributed approximately according to reported errors.  

\smallskip We also find good agreement for the four stars in common with published HIRES data (one from each of the studies by \citealt{shetrone01} and \citealt{fulbright04}, two from \citealt{cohen09}; red points in Figure \ref{fig:dra_previous1}).  For these stars, the random errors that we report for $\feh$ are about half those of the HIRES studies, despite the latter enjoying higher resolution, higher S/N and larger spectral range.  We suspect that this difference arises because we fit every pixel simultaneously while the HIRES estimates come from independent fits to individual lines.  In any case, the direct comparisons give no indication that either technique underestimates errors.  Given the different analysis techniques and range/resolution of the K10 and HIRES spectra with respect to ours, we are encouraged by the overall level of agreement displayed in Figure \ref{fig:dra_previous1}.

\section{Relations Among Physical Parameters}
\label{sec:scatterplots}
\begin{figure*}
  \includegraphics[width=6in]{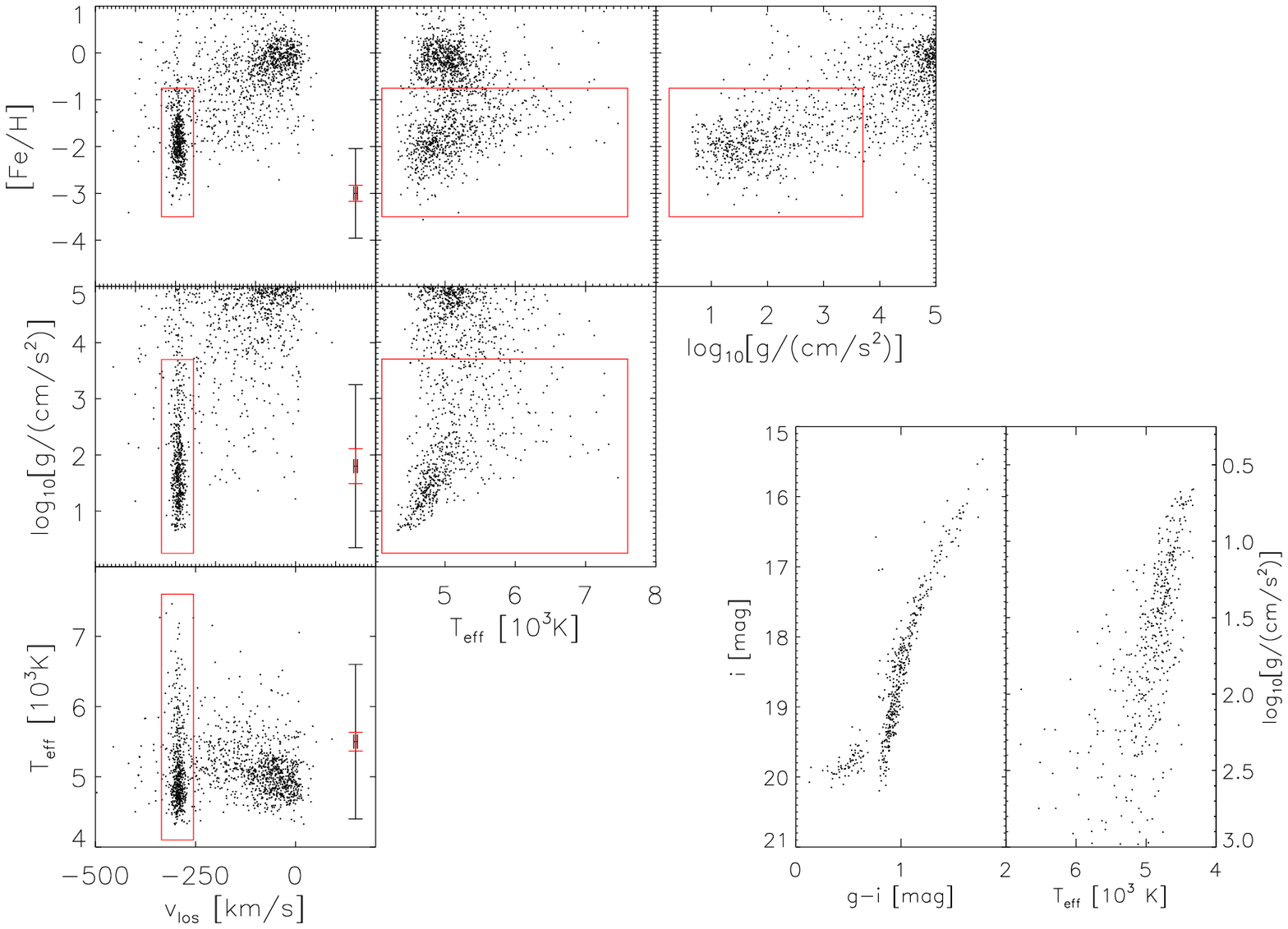}
  \caption{\scriptsize Relations among physical parameters for the \goodstars\ unique stars in our Hectochelle sample for Draco.  In the first column, error bars indicate median (red) and maximum (black) credible intervals.  Red boxes are drawn, by eye, to enclose likely Draco members with $v_{\los}\sim -290$ km s$^{-1}$ and relatively low $\logg$ and $\feh$; \ssppmembers\ stars have parameter estimates that lie inside all the boxes.  For these stars only, the two panels in the lower-right corner compare photometric and spectroscopic versions of Draco's colour-magnitude diagram. }
  \label{fig:dra_scatterplots1}
\end{figure*}

Scatterplots in Figure \ref{fig:dra_scatterplots1} show relations among physical parameters for the \goodstars\ unique stars in our sample (for stars with repeat measurements, the displayed points indicate inverse-variance-weighted means).  In most panels the distribution of points is clearly bimodal, indicating that our spectral modelling can separate Draco members from Galactic foreground in at least three dimensions: $v_{\los}$, $\logg$, $\feh$.  The Draco members cluster near $v_{\los}\sim -290$ km s$^{-1}$, while foreground interlopers have a broader velocity distribution that peaks near $v_{\los}\sim -50$ km s$^{-1}$.  As metal-poor giants, the Draco members also have $\feh$ and $\logg$ estimates that are systematically smaller than those of the observed foreground stars, which tend to be relatively metal-rich dwarfs (foreground giants are brighter than Draco's red giant branch and so are not selected for observation).  

\smallskip
Red boxes in the upper-left panels of Figure \ref{fig:dra_scatterplots1} enclose measurements that are closer to the centre of the Draco population than to the foreground population, as determined by eye.  We crudely estimate the number of Draco members in our sample by counting the number of stars with measurements that lie inside all the boxes.  We count \ssppmembers\ such likely members.  The relatively large number ($\sim 1000$) of foreground contaminants is the direct consequence of having observed at large Draco-centric radius, where members become far less numerous than foreground stars but convey important information about the influence of the external gravitational potential generated by the Milky Way.

\smallskip
The two panels in the lower-right corner of Figure \ref{fig:dra_scatterplots1} compare Draco's colour-magnitude diagram, from SDSS photometry (see also Figure \ref{fig:dra_rgbhb}), to a spectroscopic analogue based on our estimates of $\teff$ and $\logg$.  Since $\logg$ traces magnitude only for stars ascending the red giant branch, Draco's RGB is clearly separated from foreground in the spectroscopic version.  

\smallskip
Finally, the right-hand panel in Figure \ref{fig:dra_rgbhb} shows the spatial distribution of the crudely-identified likely members, tens of which lie at projected distances beyond Draco's nominal `tidal' radius, $R_t$, as estimated by \citet{ih95} from fits of \citet{king66} dynamical models to star-count data.  In agreement with \citet{munoz06}, we find likely members out to the limits of our data set, up to $\sim 3 R_t$ from Draco's centre.

\section{Summary}

We have presented a new spectroscopic data set that includes $\sim$ \ssppmembers\ Draco members distributed over an area of $\sim 4$ deg$^2$.  The observations include \goodobs\ independent measurements of line-of-sight velocity (median random error $\sigma_v\sim$ \medsigv\ km s$^{-1}$), effective temperature ($\sigma_{\teff}\sim$ \medsigteff\ K), surface gravity ($\sigma_{\logg}\sim$ \medsiglogg\ dex) and metallicity ($\sigma_{\feh}\sim$ \medsigfeh\ dex) for \goodstars\ stars, including \repeats\ independent measurements for \repeatstars\ unique stars with repeat observations.  These data will be useful for investigations ranging from galactic dynamics to chemical evolution and binary stars.  

All data from this study are publicly available.  In addition to the information provided in Tables \ref{tab:dra_table1} and \ref{tab:dra_table2}, the interested reader can download all spectra and samples from posterior PDFs (including those that do not pass quality-control criteria described in Section \ref{subsec:pdfs}) from the following web address: http://www.andrew.cmu.edu/user/mgwalker/hectochelle.  

\smallskip
We acknowledge helpful discussions with Sergey Koposov, Alan McConnachie, Jorge Pe\~narrubia, Nelson Caldwell, Evan Kirby and Alwin Mao.  We further thank Evan Kirby for making available the velocities measured from Keck/DEIMOS spectra.  We thank Andy Szentgyorgyi, Gabor Furesz and Dan Fabricant for building and maintaining Hectochelle.  We thank the MMT and SAO staff, particularly Perry Berlind, Mike Calkins, Marc Lacasse, John McAfee, Ale Milone, Ricardo Ortiz, Dennis Smith, Bill Wyatt and all the observers who participated in the Hectochelle queue.  We thank Ewan Cameron for spotting an error in Equation 12 of the original ArXiv posting.  Finally, we thank an anonymous referee for helpful suggestions.  MGW is supported by National Science Foundation grants AST-1313045, AST-1412999 and in part by National Science Foundation Grant No. PHYS-1066293 and the hospitality of the Aspen Center for Physics, where portions of this work were completed.  EO is supported by NSF grant AST-0807498 and AST-1313006.  MM is supported by NSF grants AST-0808043 and AST-1312997.  

Table \ref{tab:dra_table2} include $g$- and $i$-band magnitudes from the ninth data release of SDSS.  SDSS-III is managed by the Astrophysical Research Consortium for the Participating Institutions of the SDSS-III Collaboration including the University of Arizona, the Brazilian Participation Group, Brookhaven National Laboratory, Carnegie Mellon University, University of Florida, the French Participation Group, the German Participation Group, Harvard University, the Instituto de Astrofisica de Canarias, the Michigan State/Notre Dame/JINA Participation Group, Johns Hopkins University, Lawrence Berkeley National Laboratory, Max Planck Institute for Astrophysics, Max Planck Institute for Extraterrestrial Physics, New Mexico State University, New York University, Ohio State University, Pennsylvania State University, University of Portsmouth, Princeton University, the Spanish Participation Group, University of Tokyo, University of Utah, Vanderbilt University, University of Virginia, University of Washington, and Yale University. 


\input{astroph.bbl}
\bsp

\label{lastpage}

\end{document}

%% file: dra_ssppnewcommands1.tex
\newcommand{\obs}{$ 6107$}
\newcommand{\stars}{$ 3265$}
\newcommand{\goodobs}{$ 2813$}
\newcommand{\goodstars}{$ 1565$}
\newcommand{\repeats}{$ 1879$}
\newcommand{\repeatstars}{$  631$}
\newcommand{\maxrepeats}{$   13$}

\newcommand{\medsigv}{$ 0.88$}
\newcommand{\minsigv}{$ 0.43$}
\newcommand{\maxsigv}{$ 4.74$}
\newcommand{\medsigteff}{$  162$}
\newcommand{\minsigteff}{$   54$}
\newcommand{\maxsigteff}{$ 1100$}
\newcommand{\medsiglogg}{$ 0.37$}
\newcommand{\minsiglogg}{$ 0.10$}
\newcommand{\maxsiglogg}{$ 1.49$}
\newcommand{\medsigfeh}{$ 0.20$}
\newcommand{\minsigfeh}{$ 0.06$}
\newcommand{\maxsigfeh}{$ 0.98$}

%% file: dra_newcommands2.tex
\newcommand{\ssppdeltav}{$ 0.81$}
\newcommand{\ssppdeltateff}{$ -303$}
\newcommand{\ssppdeltalogg}{$-0.63$}
\newcommand{\ssppdeltafeh}{$-0.48$}

\newcommand{\ssppsigmav}{$ 0.42$}
\newcommand{\ssppsigmateff}{$   51$}
\newcommand{\ssppsigmalogg}{$ 0.09$}
\newcommand{\ssppsigmafeh}{$ 0.05$}

\newcommand{\coelhodeltav}{$ 0.36$}
\newcommand{\coelhodeltateff}{$  -34$}
\newcommand{\coelhodeltalogg}{$-0.06$}
\newcommand{\coelhodeltafeh}{$ 0.04$}

\newcommand{\coelhosigmav}{$ 0.43$}
\newcommand{\coelhosigmateff}{$   16$}
\newcommand{\coelhosigmalogg}{$ 0.03$}
\newcommand{\coelhosigmafeh}{$ 0.00$}

%% file: dra_newcommands3.tex
\newcommand{\ssppdeltafehkirby}{$ 0.06\pm  0.02$}

\newcommand{\ssppdeltaloggkirby}{$-0.05\pm  0.02$}

\newcommand{\ssppdeltateffkirby}{$  14\pm   10$}

\newcommand{\ssppdeltavkleyna}{$-0.01\pm  0.17$}

\newcommand{\ssppdeltavarmandroff}{$ 0.31\pm  0.23$}

\newcommand{\ssppdeltavkirby}{$ 0.02\pm  0.25$}

%% file: dra_newcommands4.tex
\newcommand{\ssppmembers}{$   468 $}

%% file: dra_log_mnras.tex
\begin{table*}
  \scriptsize
  \centering
  \begin{minipage}{100mm}
    \caption{Log of Hectochelle Observations of Draco fields}
    \begin{tabular}{@{}lllllll@{}}
      \hline
      Field&$\alpha_{2000}$$^{a}$&$\delta_{2000}$$^{a}$&UT Date&HJD$^{b}$&$N_{\rm exp}$$^{c}$&Exp. Time$^{d}$\\
      &[hh:mm:ss]&[$^{\circ}$:$\arcmin$:$\arcsec$]&[dd/mm/yyyy]&[days]&&[seconds]\\
      \hline
      Dra-01&17:20:24.65&+57:53:06.9&19/04/2006&2453844.87& 4& 7200\\
      Dra-02&17:23:50.00&+57:52:12.0&25/04/2006&2453850.78& 5& 4846\\
      Dra-03&17:20:24.64&+57:53:06.9&23/02/2007&2454154.98& 3& 5400\\
      Dra-04&17:17:37.81&+57:46:30.2&27/02/2007&2454158.96& 3& 5400\\
      &&&11/03/2007&2454170.89& 4& 7200\\
      Dra-05&17:20:38.90&+57:28:04.3&03/03/2007&2454162.94& 3& 5400\\
      &&&09/03/2007&2454168.90& 3& 5400\\
      Dra-06&17:19:23.67&+58:28:22.4&06/03/2007&2454165.90& 3& 5400\\
      Dra-07&17:14:24.40&+57:28:47.6&22/04/2007&2454212.89& 3& 5400\\
      Dra-08&17:26:35.35&+58:15:25.6&23/04/2007&2454213.83& 3& 5400\\
      Dra-09&17:30:06.97&+57:38:24.2&24/02/2008&2454520.94& 5& 6000\\
      Dra-10&17:20:25.01&+57:53:11.7&27/02/2008&2454523.99& 3& 3600\\
      Dra-11&17:11:57.01&+58:18:03.1&27/02/2008&2454523.93& 3& 4500\\
      Dra-12&17:10:09.01&+57:29:41.0&20/03/2009&2454910.85& 2& 4096\\
      &&&21/03/2009&2454911.86& 3& 7200\\
      Dra-13&17:20:06.72&+57:55:32.6&21/03/2009&2454911.96& 2& 4800\\
      &&&24/03/2009&2454914.92& 4& 7200\\
      Dra-14&17:41:60.00&+56:00:00.0&23/03/2009&2454913.87& 4& 9600\\
      Dra-15&17:20:15.99&+57:55:30.0&15/05/2010&2455331.74& 4& 6000\\
      Dra-16&17:20:15.99&+57:55:30.0&15/05/2010&2455331.84& 3& 4500\\
      Dra-17&17:15:23.52&+57:55:42.0&16/05/2010&2455332.73& 2& 3600\\
      &&&16/05/2010&2455332.78& 3& 4500\\
      Dra-18&17:20:23.58&+58:24:30.0&16/05/2010&2455332.85& 3& 4500\\
      Dra-19&17:20:11.57&+57:24:54.0&16/05/2010&2455332.91& 3& 4500\\
      Dra-20&17:20:15.99&+57:55:30.0&29/01/2011&2455590.99& 3& 3600\\
      &&&25/05/2011&2455706.92& 4& 4500\\
      Dra-21&17:20:15.99&+57:55:30.0&25/05/2011&2455706.77& 3& 5400\\
      Dra-22&17:20:15.99&+57:55:30.0&25/05/2011&2455706.84& 3& 5400\\
      Dra-23&17:22:55.93&+58:02:34.0&26/05/2011&2455707.75& 3& 5400\\
      Dra-24&17:17:39.05&+57:53:14.0&26/05/2011&2455707.83& 3& 5400\\
      Dra-25&17:20:46.45&+57:17:29.3&26/05/2011&2455707.91& 4& 4800\\
      Dra-26&17:20:00.58&+58:37:30.6&27/05/2011&2455708.86& 3& 5400\\
      Dra-27&17:20:15.99&+57:55:30.0&27/05/2011&2455708.93& 3& 3600\\
      &&&31/05/2011&2455712.87& 3& 3600\\
      Dra-28&17:20:19.72&+57:52:21.9&27/05/2011&2455708.76& 4& 9600\\
      &&&28/05/2011&2455709.89& 3& 6000\\
      Dra-29&17:19:02.64&+57:22:36.0&29/05/2011&2455710.84& 3& 5400\\
      Dra-30&17:22:06.84&+58:27:53.3&29/05/2011&2455710.91& 3& 4500\\
      Dra-31&17:20:17.49&+57:54:42.0&29/05/2011&2455710.76& 3& 5400\\
      Dra-32&17:26:54.32&+58:29:28.6&30/05/2011&2455711.90& 3& 5400\\
      &&&31/05/2011&2455712.93& 3& 3600\\
      Dra-33&17:14:13.51&+57:40:52.9&31/05/2011&2455712.78& 4& 7200\\
      \hline
    \end{tabular}
    \\
    $^{a}$ central coordinates of field\\
    $^{b}$heliocentric Julian date at beginning of first sub-exposure\\
    $^{c}$number of sub-exposures\\
    $^{d}$exposure time, summed over all sub-exposures\\
  \end{minipage}
  \label{tab:log}
\end{table*}